\documentclass{aa}
\usepackage{txfonts}
\usepackage{graphicx}

\usepackage{natbib}
\bibpunct{(}{)}{;}{a}{}{,} 

\begin{document}

\title{ Accretion and ejection properties of embedded protostars: the case of HH26, HH34 and HH46 IRS
	\thanks{Based on observations collected at the European Southern Observatory, Chile (ESO Programmes 072.C-0176(A), 074.C-0235(A)) }}	

\author{	S. Antoniucci \inst{1,2},
	B. Nisini  \inst{2},
	T. Giannini \inst{2}
	\and D. Lorenzetti \inst{2}
	}

\institute{	Universit\`a degli Studi di Roma `Tor Vergata', via della Ricerca Scientifica 1, I-00133 Roma, Italy \and 
		INAF-Osservatorio Astronomico di Roma, Via di Frascati 33, I-00040 Monteporzio Catone, Italy  }	

\offprints{Simone Antoniucci, email address: antoniucci@oa-roma.inaf.it}

\date{Received date / Accepted date}

\abstract{}
{We present the results of a spectroscopic analysis on three young embedded sources  (HH26 IRS, HH34 IRS and HH46 IRS) belonging to different star-forming regions and displaying well developed jet structures. The aim is to investigate the source accretion and ejection properties and their connection.}
{We used VLT-ISAAC near-IR medium resolution ($R\sim9000$) spectra ($H$ and $K$ bands) to derive, in a
self-consistent way,  parameters like the star luminosity, the accretion luminosity and the mass accretion rate. Mass ejection rates have also been estimated from the analysis of different emission features.}
{The spectra present several emission lines but no photospheric features in absorption, indicating a large veiling in both $H$ and $K$ bands. In addition to features commonly observed in jet driving sources ([\ion{Fe}{ii}], H$_2$, \ion{H}{i}, CO), we detect a number of emission lines due to permitted atomic transitions, such as \ion{Na}{i} and \ion{Ti}{i} that are only 2-5 times weaker than the Br$\gamma$ line. 
Some of these features remain unidentified. 
Emission from \ion{Na}{i} 2.2$\mu$m doublet is observed along with CO(2-0) band-head emission, indicating a common origin in an inner gaseous disc heated by accretion.
We find that accretion provides about 50\% and 80\% of the bolometric luminosity in HH26 IRS and HH34 IRS, as expected for accreting young objects. Mass accretion and loss rates spanning $10^{-6}$--$10^{-8}$ M$_{\sun}$\,yr$^{-1}$ have been measured. The derived $\dot{M}_\mathrm{loss}/\dot{M}_\mathrm{acc}$ is $\sim$0.01 for HH26 IRS and HH34 IRS, and $>$0.1 for HH46 IRS.  These numbers are in the range of values predicted by MHD jet launching models and found in the most active classical T Tauri stars.}
{Comparison with other spectroscopic studies performed on Class Is seems to indicate that Class Is actually having accretion-dominated luminosities are a limited number. Although the analysed sample is small, we can tentatively define some criteria to characterise such sources: they have $K$-band veiling larger than 2 and in the majority of the cases present IR features of CO and \ion{Na}{i} in emission, although these do not directly correlate with the accretion luminosity. Class Is with massive jets have high $L_{\mathrm{acc}}/L_{\mathrm{bol}}$ ratios but 
not all the identified accretion-dominated objects present a jet. As suggested by the SEDs of our three objects, the accretion-dominated objects could be in an evolutionary transition phase between Class 0 and I.
Studies of the kind presented here but on larger samples of possible candidates should be performed in order to test and refine these criteria.}

\keywords{ Stars: formation -- Stars: evolution -- Infrared: stars -- ISM: jets and outflows}

\titlerunning{Accretion and ejection in embedded protostars.}
\authorrunning{Antoniucci et al.}
\maketitle

\section{Introduction}

The process of mass accretion accompanying the formation of solar type stars is always associated with mass ejection in form of collimated
jets, that extend from few AU up to parsecs distance from the exciting source. According to an established class of models \citep{koenigl00,casse00},
accretion and ejection are indeed intimately related through the presence of a magnetised accretion disc: the jets carry away the excess  
angular momentum, so that part of the  disc material can move toward 
the star. The efficiency of this process is measured by the ratio between the mass ejection and mass accretion rates, and depends on the  jet acceleration mechanism at work. Measurements of such an efficiency have been so far obtained only for classical  T Tauri stars, whose accretion properties are rather well studied through the observation of the excess continuum emission at optical and UV wavelengths 
Values of $\dot{M}_\mathrm{loss}/\dot{M}_\mathrm{acc}$ in the range 1-10\% have been found by different studies \citep[e.g.][]{hartigan95,woitas05,ferreira06}.
It is nevertheless important to test accretion/ejection models in young sources at earlier stages of evolution, when  accretion 
dominates the energetics of the system and thus the mechanism to extract angular momentum is expected to be more efficient.
To this aim, an interesting sample of objects  are Class I sources, i.e. the class of embedded stars characterised by a steeply 
rising IR spectral energy distribution (SED) between 2 and 10 $\mu$m and usually considered younger than visible T Tauri stars (the Class II sources).
However, the high extinction pertaining to these objects strongly limits the measurement of their stellar and accretion properties, 
needed to prove that they are indeed in a phase of higher accretion with respect to Class II sources.
The general assumption so far applied has been that most of the bolometric luminosity of Class I objects is due to accretion. 

Recently, however, thanks to the use of high dispersion sensitive instrumentation, it has become possible to define the stellar properties
of small samples of Class I stars through their weak photospheric lines detected in the optical scattered light \citep{white04}
and in the near-IR direct emission \citep{greene02,nisini05a,doppmann05}. Such studies have shown that the characteristics 
of Class I objects vary in fact significantly. In particular, the accretion luminosity may span from few percent up to 80\% of the bolometric luminosity; these findings show that  not all sources defined as Class I are indeed actively accreting objects and suggest that 
a classification based on different criteria is indeed required.

\citet{white04}, notably, derived from their analysis of the scattered light spectra of Taurus-Auriga sources, that the average $\dot{M}_\mathrm{loss}/\dot{M}_\mathrm{acc}$ for Class I of their sample is larger than for Class II and close to unity. They interpret this result as due to an observational bias induced by the effect of disc orientation in their optical spectra; if the Class I are seen prevalently edge-on,  then the extended region emitting the forbidden lines from which they estimate $\dot{M}_\mathrm{loss}$ is seen more directly than the obscured stellar photosphere. It is clear that such kind of biases can be minimised by performing spectroscopic observations directly in the IR, where features originating in the photosphere, in the accretion region and in the jet can be simultaneously detected by instrumentation which is sensitive enough.

In this framework, we report here  the results of near-IR spectroscopic observations at medium resolution of three embedded sources (HH34 IRS, HH26 IRS and HH46 IRS)  displaying well developed jet structures and having a spectral index between 2 and 10\,$\mu$m, typical of Class I objects. 
We have derived accretion and ejection parameters of these sources through the analysis 
of the different features detected on the spectra. The goal is to study how much of their energy is due to accretion and to investigate the efficiency of the ejection mechanism.

We describe the sample and the observations in Sec. 2 and report the results in Sec. 3; in Sec. 4 we present the procedures applied for the analysis of the data and infer the physical properties of the objects and their jets. These results are then discussed in Sec. 5, where a comparison with similar sources analysed in previous studies will be also made. Main conclusions of our work are summarised in Sec. 6.

\begin{table*}[!t]

\begin{center}
\caption[]{The observed targets and their main observational properties.}
\begin{tabular}{l c c | c | c c | c c | c c c c}
\hline
\hline
Source         & R.A.(2000)			&DEC.(2000)		&$m_\mathrm{J}$& \multicolumn{2}{|c|}{$m_\mathrm{H}$}	 	& \multicolumn{2}{c|}{$m_\mathrm{K}$}	& $D$		& $\alpha^{(a)}$		& $L_\mathrm{bol}$		\\
					&						&				&(mag)					&\multicolumn{2}{|c|}{(mag)}&\multicolumn{2}{|c|}{(mag)}&(pc)	&	&	(L$_{\sun}$)	\\
					&						&				&2MASS$^{(1)}$&2MASS$^{(1)}$&This work$^{(1)}$&2MASS$^{(b)}$&This work$^{(b)}$&	&	&		\\
\hline
HH 26 IRS	& 05 46 03.9	& -00 14 52			&16.77& 14.07 &14.6  & 11.88 & 12.3 		& 450 		& 2.01 			& 4.6--9.2	\\	
HH 34 IRS	& 05 35 29.9	& -06 26 58	 		&15.06& 13.60 &13.5 	& 12.38 & 12.4 		& 460 		& 1.14 			& 12.4--19.9	 \\	
HH 46 IRS	& 08 25 43.9	& -51 00 36	 		&14.20& 12.88 &14.8 	& 12.72 & 13.4 		& 450 		& 1.96 			& $<$15.0$^{(c)}$ \\
\hline 
\end{tabular}
\label{targets}
\end{center}
\vspace{0.2 cm}
\small{Notes. (a) The spectral index $\alpha=d\mathrm{Log}(\lambda F_{\lambda})/d\mathrm{Log}(\lambda)$ is calculated between 2 and 10 $\mu$m. (b) The magnitude values in the $H$ and $J$ bands have been estimated from the calibrated spectra. (c) Total luminosity: the source is a binary (see text for details). References. (1) from the 2MASS catalogue \citep{2mass}.}

\end{table*}

\section{Description of the sample and observations}

\subsection{The sample}
Our sample consists of three sources, namely HH26 IRS, HH34 IRS and HH46 IRS, 
selected among sources having a positive spectral index between 2 and 10 $\mu$m 
and showing well defined jet structures observed in the near-IR very close to the source itself.    
The basic observational properties of the targets are listed in Tab.\ref{targets}.

\textbf{HH34 IRS} (located in the L1641 cloud, at a distance of 460 pc) is the exciting source of the spectacular parsec-scale HH34 flow \citep{bally94}. 
It was initially detected as a faint optical star by \citet{reipurth86}, but \citet{reipurth00} showed that the optical emission is coming 
from a very compact reflection nebulosity, while the source itself is an embedded IR object displaced at about 0.1 arcsec from this 
nebulosity. HH34 IRS has been detected  by IRAS and at sub-mm/mm wavelengths by \citet{reipurth93} and 
\citet{johnstone06a}, who show that a dense cloud core surrounds the object. The HH34 jet has been studied in 
detail both in the optical \citep{reipurth02,bacciotti99} and in the IR \citep{podio06,takami06,garcia_lopez07}, where 
the H$_2$ molecular counterpart of the atomic jet has been also detected. Jet proper motion analysis has shown that the jet is oriented 
with an angle ranging from 20 to 30$^\mathrm{o}$ to the plane of the sky \citep{eisloeffel92,heathcote92}. 

\textbf{HH26 IRS} (located in the L1630 molecular cloud at a distance of 450 pc) is an embedded object, recognised to be the exciting source 
of the HH26A/C/D chain  and of the associated molecular outflow by \citet{davis97}. \citet{davis02} discovered that this 
source is also driving a small-scale H$_2$ jet with a length of a few arcsec. An IR spectrum of HH26 IR from 2 to 4 $\mu$m has been obtained by 
\citet{simon04}, showing emission line features, from H$_2$, HI and CO, typical of active young sources.
Observations at sub-mm/mm wavelengths of this object have been obtained  by \citet{lis99} and \citet{johnstone01}.

\textbf{HH46 IRS} is located in an isolated Bok globule close to the Gum nebula (at a distance of 450 pc). The object actually consists of an 
embedded young binary system of 0.26\arcsec separation \citep{reipurth00} that illuminates a reflection nebula from which the HH46/47 
system of Herbig Haro objects emerges.
At infrared wavelengths, the redshifted counter-jet has been detected penetrating through the reflection nebula down to the exciting source 
\citep{eisloeffel94a}. HH46-IRS is associated with the IRAS object IRAS08242-5050 and has been detected at millimetre by \citet{reipurth93} . A spectrum at mid-IR wavelength of HH46-IRS has been recently obtained by Spitzer \citep{noriega-crespo04}. Proper 
motion measurements of the HH46/47 objects indicate the HH46 IRS jet is inclined by $\sim$34$^\mathrm{o}$ to the plane of the sky.
\citep{eisloeffel94b}.

\subsection{Observations}
The observations were performed using the VLT-UT1 ISAAC spectrograph in SW mode on 28, 29 and 30 December 2004.
Three spectral regions have been investigated: one in the $H$ band (1.57-1.65 $\mu$m for HH34 IRS and HH46 IRS and 1.61-1.69 $\mu$m for HH26 IRS), and two contiguous segments in 
the $K$ band (2.08-2.30 $\mu$m). These spectral regions have been chosen because here lie 
important emission lines tracing accretion and ejection (e.g. \ion{H}{i} and [\ion{Fe}{ii}] ), and also because of the presence of several 
diagnostic photospheric absorption features which, if detected, are suitable for spectral classification of the forming star.

All spectra were obtained using a 0.3\arcsec slit, providing a final nominal spectral resolution of about 10000 in 
the $H$ band and 8900 in the $K$ band. The slit has been aligned parallel to the jet. All the work on the raw images has been performed using the IRAF software package, following the standard procedures for bad pixel removal, flat-fielding and sky subtraction. 
Spectra of standard stars have been acquired at airmasses similar to those of the scientific spectra and used, after removal of any intrinsic line, to correct for the telluric absorptions and to calibrate the flux scale.

 Observations of science targets and standards were performed with similar seeing conditions ($\sim$0.6\arcsec for HH34 and HH46 IRS and $\sim$1\arcsec for HH26 IRS), so that flux losses are assumed to be comparable in both cases for point-like objects and no correction has been applied to the spectra of the sources (for jets see Sec. \ref{jmf}).

From the calibrated spectra we derived an estimate of the magnitudes in the $H$ and $K$ bands, which we list in Tab. \ref{targets} together with the 2MASS magnitudes. The agreement is very good for HH34 IRS while for the other sources the differences are $<$0.6 mag except for the $H$ magnitude of HH46 IRS; in this case, however, the 2MASS $H$ magnitude appears to be strangely high, being almost equal to the $K$ magnitude, providing a $J-H$ colour unusual for this kind of sources.

Wavelength calibration has been performed using the atmospheric OH emission lines whenever it was possible, 
or, alternately, using Xenon/Argon lamp lines, always refining the alignment against known atmospheric absorption 
features. The described procedure leads to a calibration error that we estimate of the order of 0.1 \AA \, for the $H$ band and 
the first $K$ band segments, and 0.2 \AA \, for the other $K$ band spectral region investigated. 

Velocity information on the observed lines ($V_\mathrm{LSR}$ and $\Delta V$) has been obtained through a Gaussian fit. 
The uncertainty on the line widths is of about 10 km s$^{-1}$, as we estimate considering the typical difference between 
the fitted FWHM and the one directly measured on the spectra.
Concerning radial velocities, the contribution $V_\mathrm{cloud}$ of the star-forming clouds hosting the sources, i.e. 11 km s$^{-1}$ for HH26 IR \citep{gibb95}, 8.5 km s$^{-1}$ for HH34 IRS \citep{anglada95} and 8 km s$^{-1}$ for HH46 IRS in Vela \citep{chernin91}, has been subtracted from the measured $V_\mathrm{LSR}$.

\section{Results}

\subsection{Observed spectral features}

\begin{figure}[!b]
\caption{The $H$ and $K$ band normalised spectra of HH26 IRS.} 
\includegraphics[width=8.9cm]{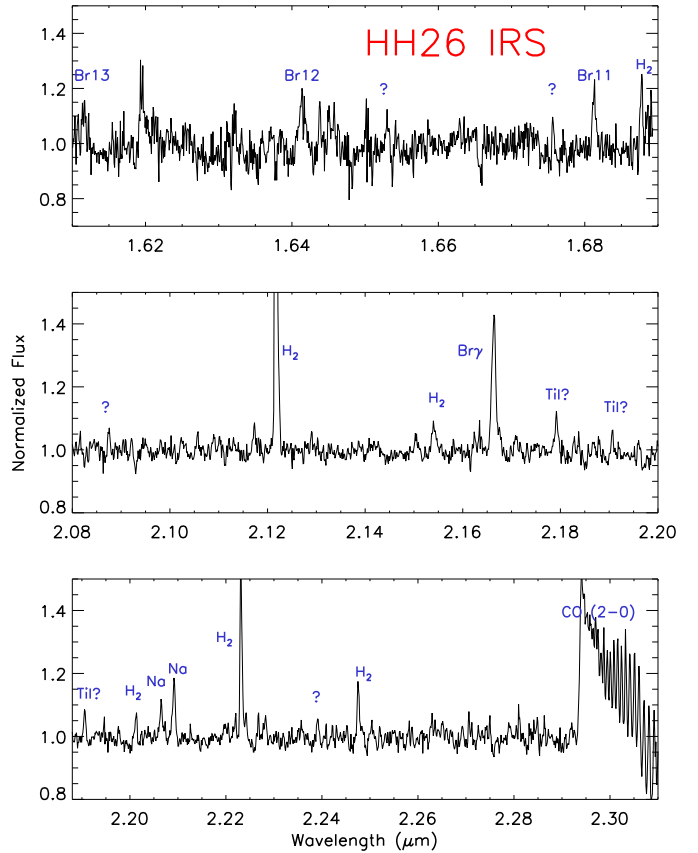}\\
\label{plots26} 
\end{figure}

\begin{figure}[!t]
\caption{The $H$ and $K$ band normalised spectra of HH34 IRS.} 
\includegraphics[width=8.9cm]{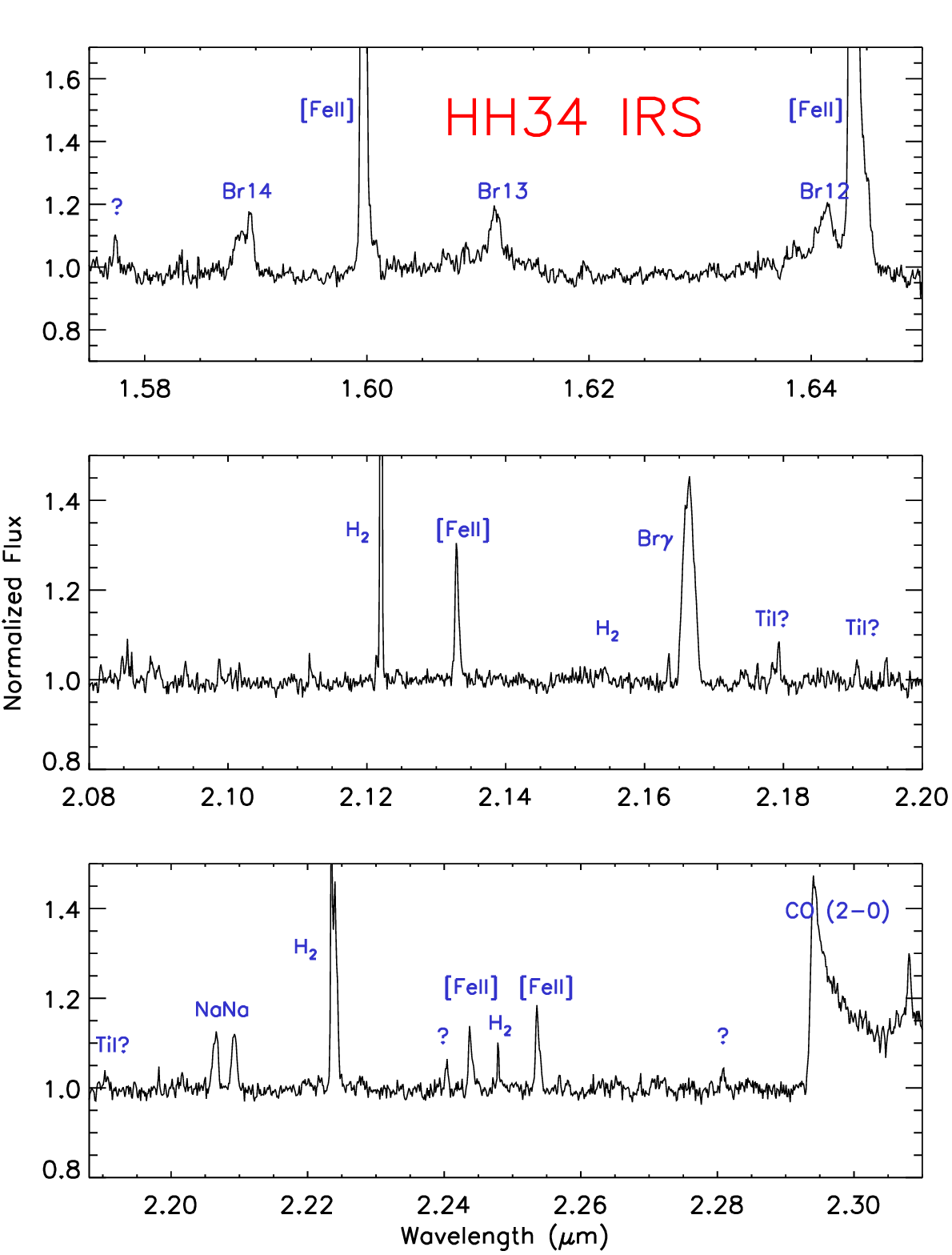}\\
\label{plots34} 
\end{figure}

\begin{figure}[!t]
\caption{The $H$ and $K$ band normalised spectra of HH46 IRS.} 
\includegraphics[width=8.9cm]{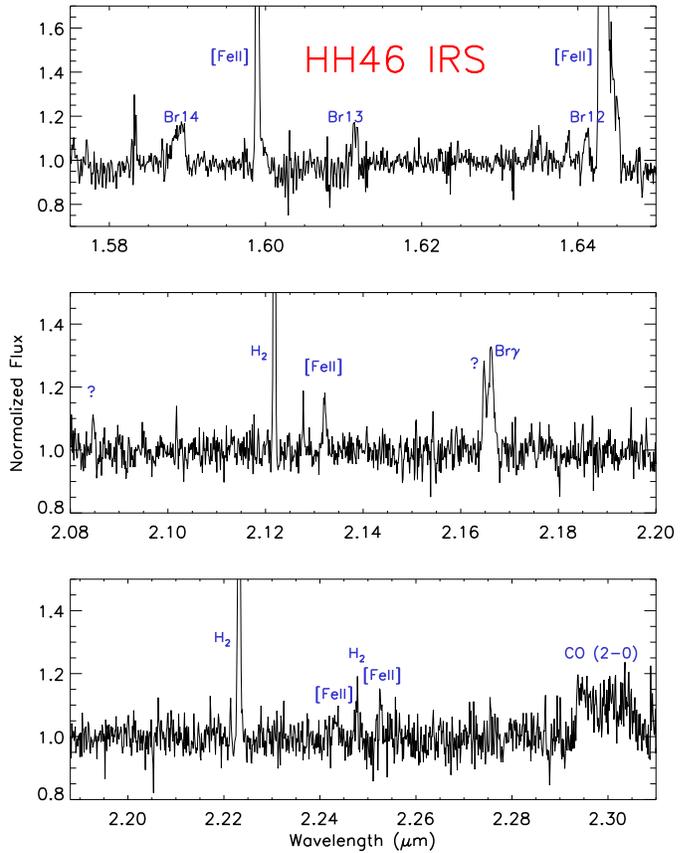}
\label{plots46} 
\end{figure}

The continuum-normalised spectra of the three sources are shown in Fig. \ref{plots26}, \ref{plots34} and \ref{plots46}.
The spectra are characterised by the presence of several emission lines, while there seems to be no evidence of absorption 
features from the central object photosphere.
Depending on the rms noise measured on the spectra, we estimate upper limits on the 
line/continuum ratio of the undetected photospheric lines between $\sim$ 5-30 \%. 
These ratios give an indication of the minimum absorption line equivalent width measurable for a given spectrum. Hence, a lower limit to the 
veiling can be inferred using the relationship $r=EW/EW'-1$, connecting the veiling $r$ to the intrinsic ($EW$) and measured ($EW'$) equivalent 
width of a line. 
Assuming a late K spectral type for the sources, being typical intrinsic equivalent widths around 1.5\AA \, for the strongest absorption 
features in the $K$ band in this type of stars \citep[e.g.][]{wallace97}, one derives veilings $r_K \gtrsim 1, 2, 5$ for HH46 IRS, HH26 IRS and HH34 IRS, respectively.

Among the emission lines, there are several H$_2$ and [\ion{Fe}{ii}] transitions coming from the jet, as testified by the fact that these 
lines are not confined to the source but appear extended in the jet direction in the spectral images. [\ion{Fe}{ii}] lines trace regions with moderate ionisation, temperatures between 8000 and 15000 K and electron densities up to 10$^5$ cm$^{-3}$, 
which are commonly found in the initial part of a jet, close to the source. Rovibrational H$_2$, on the other hand, is excited in a molecular gas at temperatures of the order of 2000--3000 K and total densities of $\sim$10$^4$-10$^5$ cm$^{-3}$. 
While HH34 IRS and HH46 IRS show both [\ion{Fe}{ii}] and H$_2$ lines, in HH26 IRS only these latter are detected. 
This could be an indication that the jet has low excitation conditions or it is composed mostly of molecular material. 
Similar molecular jets have been found in a small number of Class I sources \citep{davis01}. 

A full account of the jet physical properties derivable from the detected emission lines will be given elsewhere \citep[][in preparation]{garcia_lopez07,calzoletti07}.

\begin{figure}[!b]
\caption{\ion{H}{i} line profiles for (from left to rigth) HH26 IRS, HH34 IRS and HH46 IRS.} 
\centering
\includegraphics[width=2.9cm]{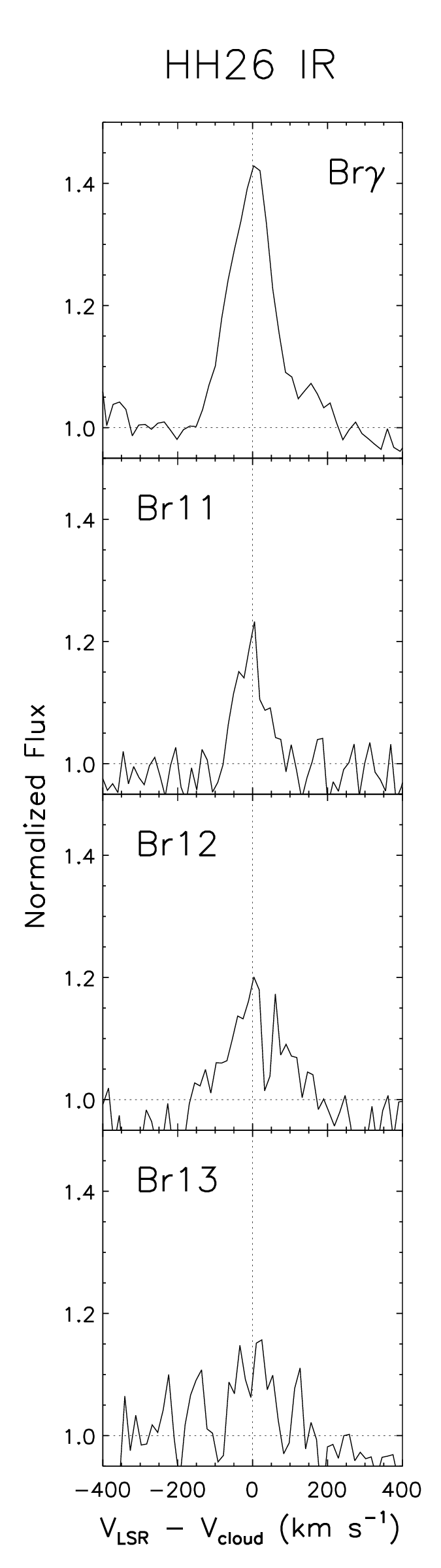}
\includegraphics[width=2.9cm]{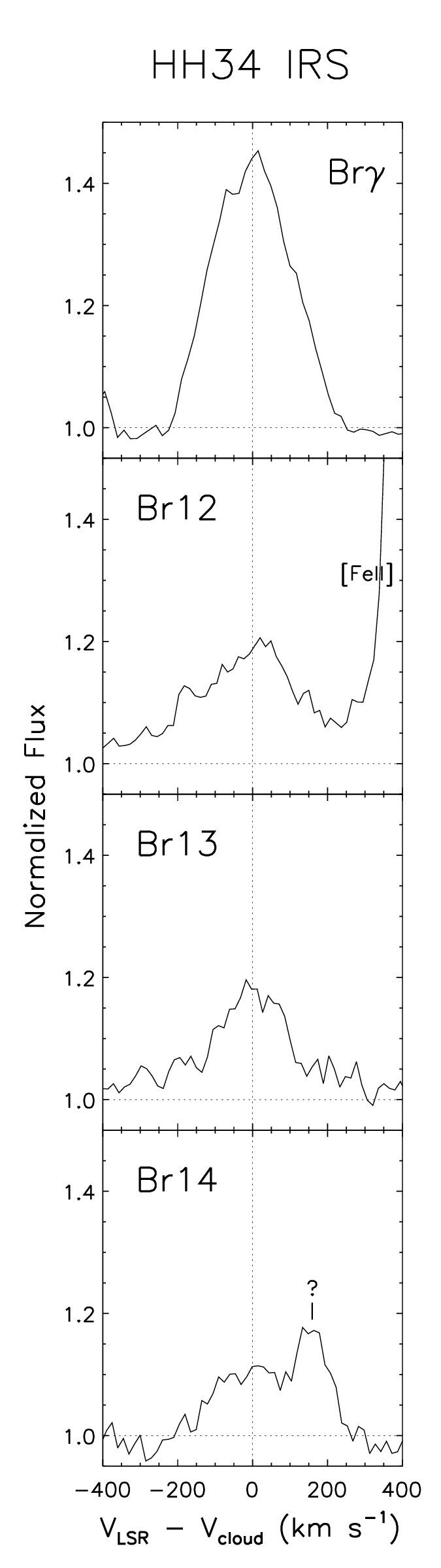}
\includegraphics[width=2.9cm]{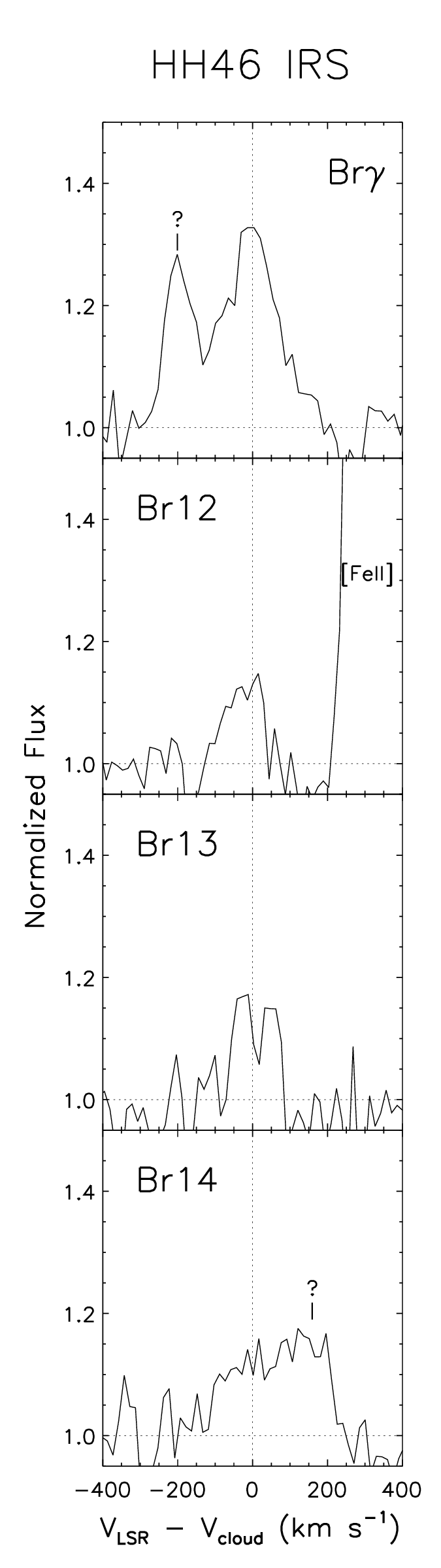}
%
\label{fig_lines} 
\end{figure}

\begin{figure}[!h]
\caption{\ion{Na}{i} line profiles for (from left to rigth) HH26 IRS and HH34 IRS.} 
%
\centering
\includegraphics[width=3.2cm]{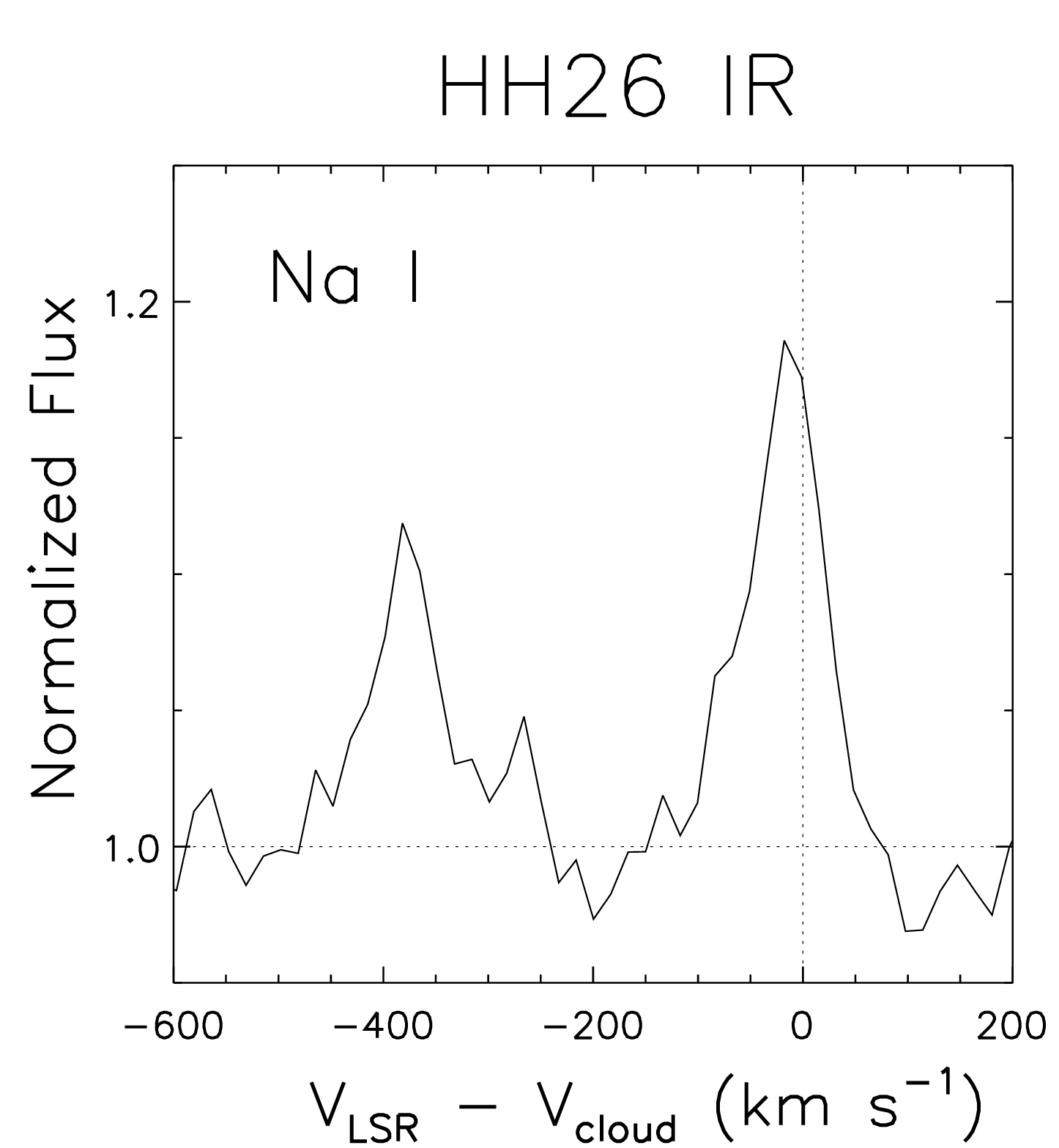}
\hspace{.5cm}
\includegraphics[width=3.2cm]{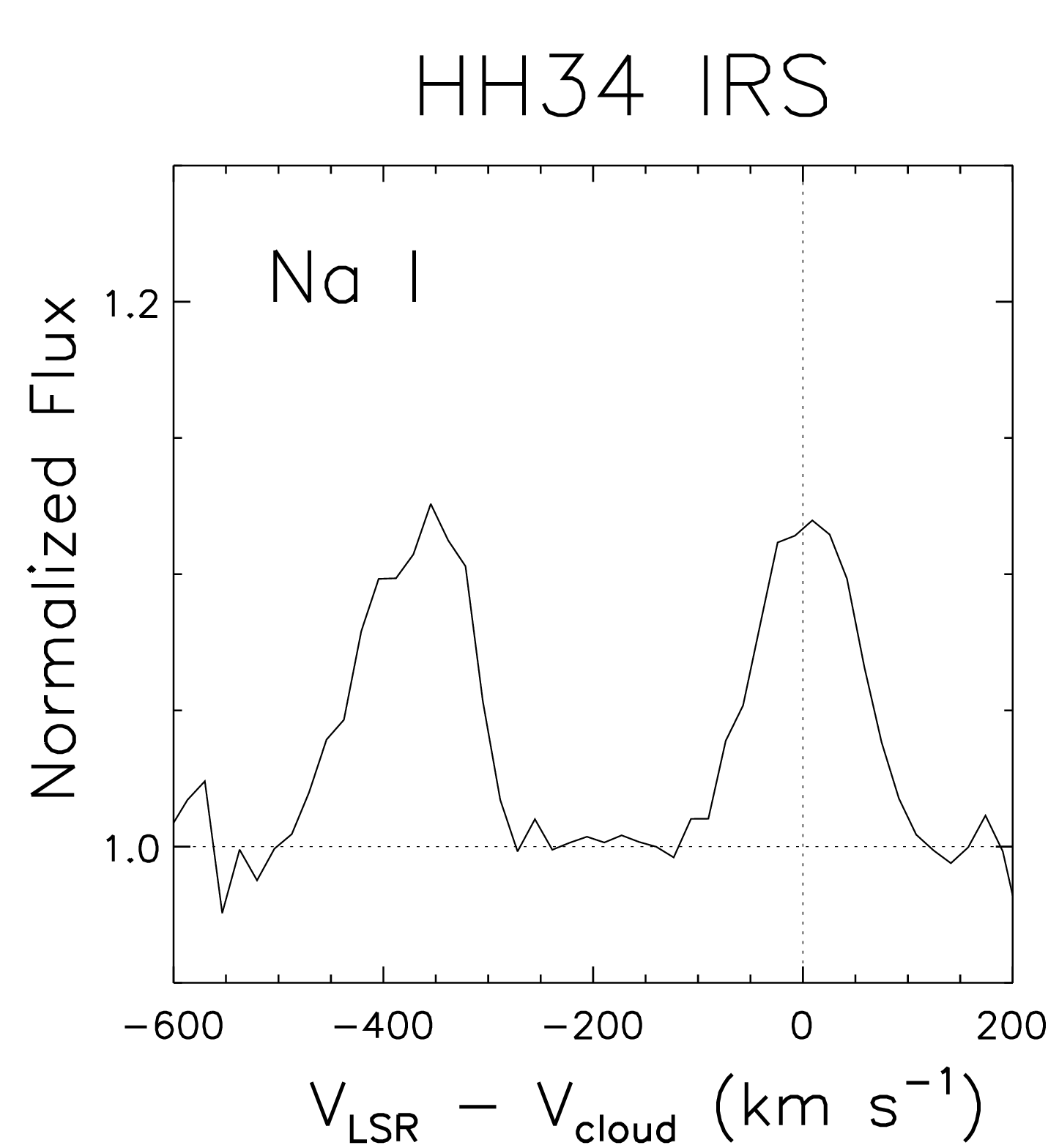}
%
\label{fig_lines_na} 
\end{figure}

Different permitted emission lines are also detected, the most prominent being the \ion{H}{i} recombination lines of the Brackett series  
(see Fig. \ref{fig_lines}) and the \ion{Na}{i} doublet lines around 2.20 $\mu$m (Fig. \ref{fig_lines_na}), which are observed in HH34 IRS and HH26 IRS. 
Such lines, whose emission is confined toward the source, trace high density gas excited either in the accretion region 
or at the base of stellar/disc winds. 
The  \ion{H}{i} lines have large widths as is usual in these kind of sources \citep[e.g.][]{nisini05a}; the Br$\gamma$ line in HH34 IRS has a width of order 240 km\,s$^{-1}$ while in the other sources it is of order 150 km\,s$^{-1}$. The peak of the emission is just slightly redshifted, but the profile is not always symmetric, as it extends more in the blue in HH26 IRS.
In the case of HH46 IRS the line profile is difficult to interpret since the line is merged with another prominent emission feature whose identification is not certain, although the wavelength (2.165 $\mu$m) is compatible with \ion{He}{i}. The velocity shift of about -200  km\,s$^{-1}$ (similar to the one measured on \ion{Fe}{ii}) could indicate a Br$\gamma$ high velocity component associated with the \ion{Fe}{ii} jet but confined on the source and not extended along the jet axis. Such a high velocity component is however not detected in the other Brackett series lines.

Br$\gamma$ equivalent widths measured on the sources span from -4.8 to -8.9 \AA (see Tab. \ref{param2}).

Band-head emission from CO is also detected in the sources, even though the emission is much weaker in HH46 IRS; this is a tracer for dense molecular gas at T $\sim$2000-3000 K, usually interpreted as a sign of the presence of either an active accretion disc or of a neutral wind \citep[e.g.][]{carr89,carr93,najita96}.

Finally, we detect other lines for which we provide tentative identifications, namely \ion{Ti}{i} lines at 2.179 and 2.190 $\mu$m 
(visible in HH26 IRS and also in HH34 IRS) and other features that remain unidentified (e.g. at 2.239 $\mu$m in HH26 IRS and at 2.240, 2.281 $\mu$m in HH34 IRS).
The \ion{Ti}{i} lines (that are well detected especially HH26 IRS) are broader than H$_2$ lines, while they basically have the same width as the \ion{Na}{i} doublet lines (around 80 km\,s$^{-1}$), thus suggesting that they probably originate in the same region around the central source.  

A list of all the observed emission features having S/N greater than 3 and line-widths at least equal to the instrumental resolution is reported in Tabs \ref{lineshh26}, \ref{lineshh34} and \ref{lineshh46}, together with their fluxes and tentative identifications based on the  NIST\footnote{http://physics.nist.gov/PhysRefData/ASD1} and Atomic Line List\footnote{http://www.pa.uky.edu/~peter/atomic databases}.

Radial velocity information for the brightest lines of the detected series (\ion{H}{i} Br$\gamma$, H$_2$ at 2.12 $\mu$m, \ion{Fe}{ii} at 1.64 $\mu$m and \ion{Na}{i} at 2.2 $\mu$m) is given in Tab. \ref{t-vel}.

\begin{figure}[!th]
\caption{\label{f-sed} \small{Spectral Energy Distributions of the three sources, derived from the values reported in  Tab. \ref{t-sed}. Points from different instruments are displayed with different symbols (2MASS: crosses, MSX: triangles, IRAS: squares, SPITZER: diamonds, far IR/mm measurements: stars.}}
\vspace{.3cm}
\centering
\includegraphics[width=6.3cm]{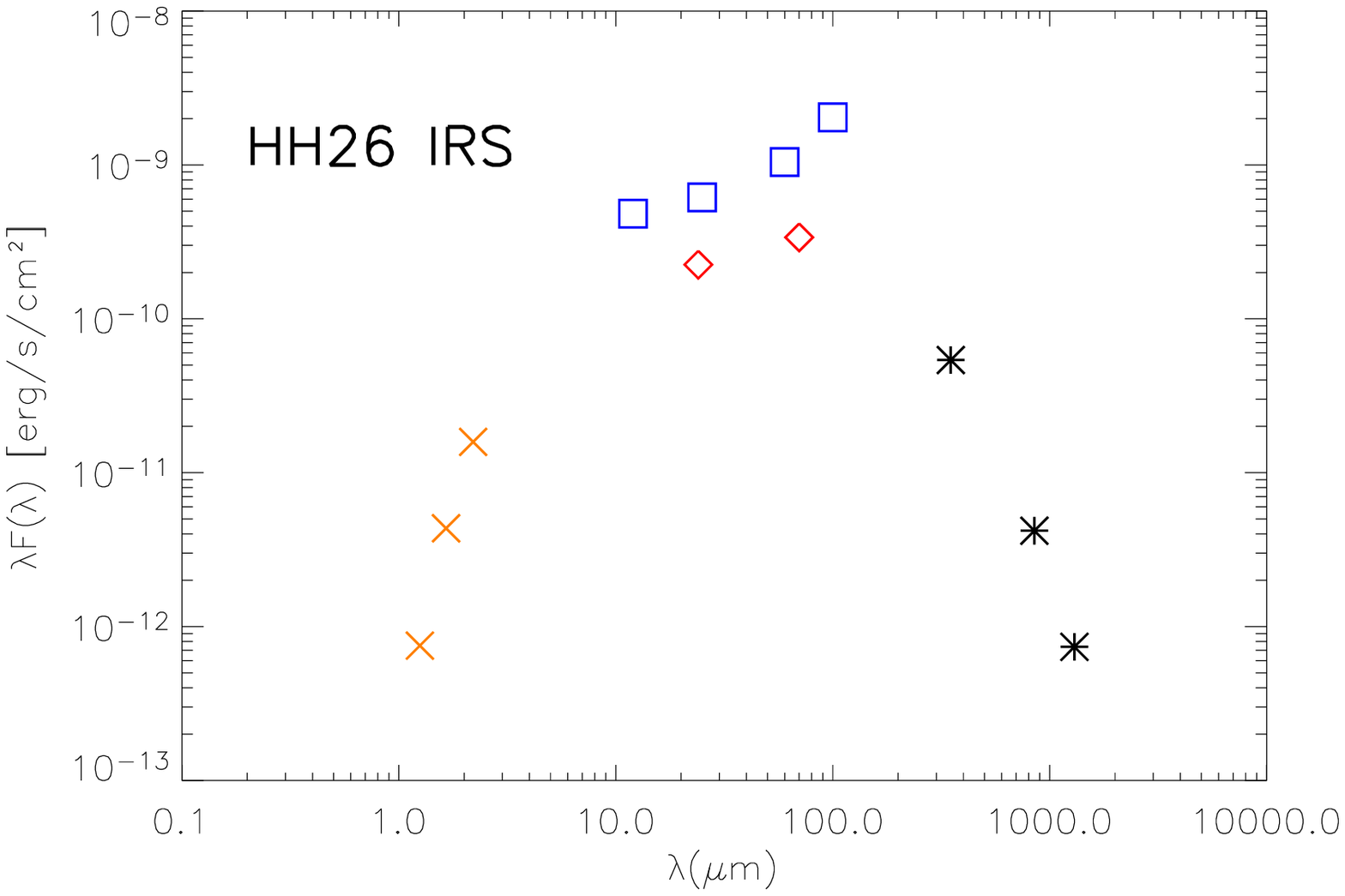}\\
\includegraphics[width=6.3cm]{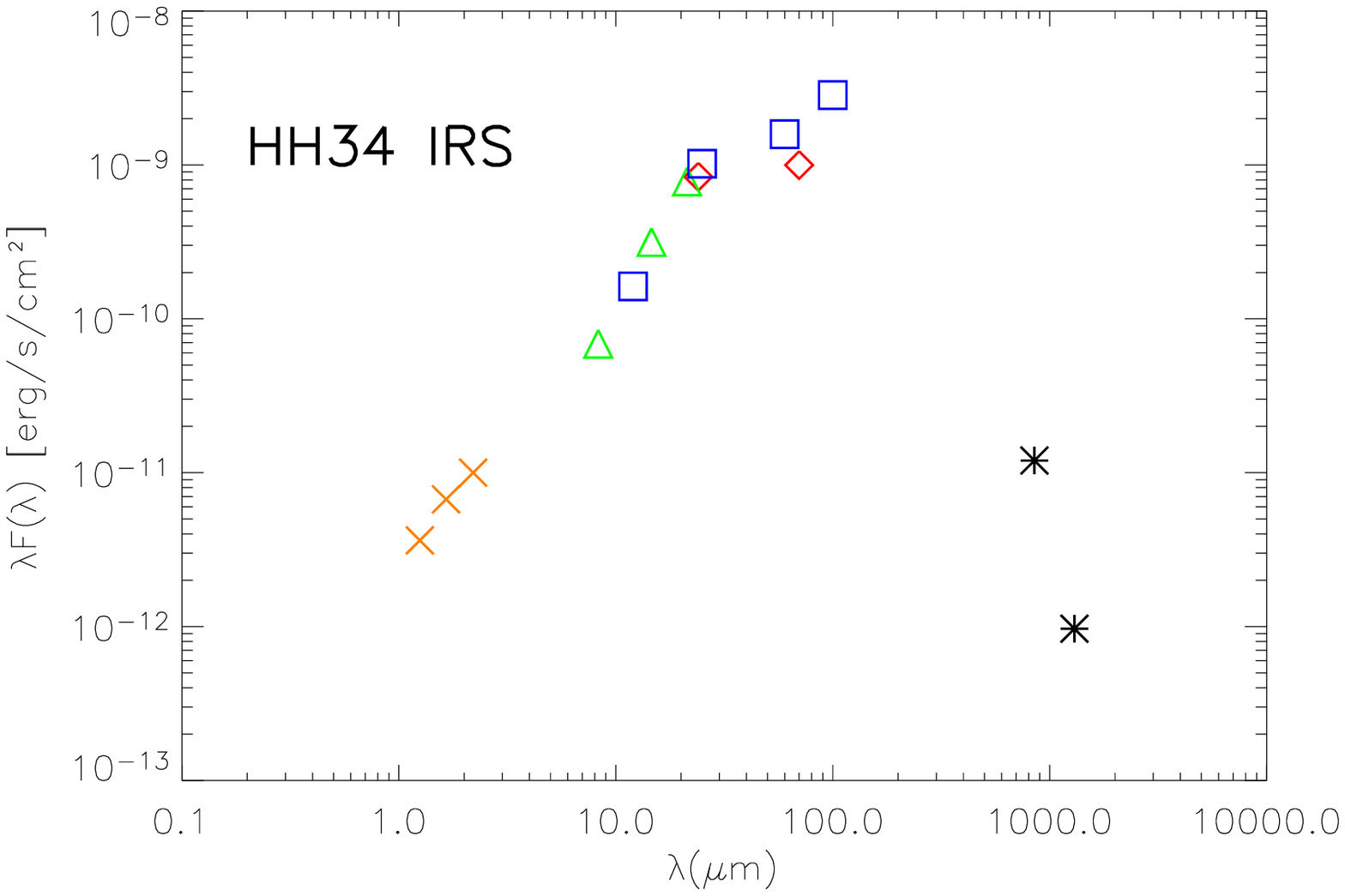}\\
\includegraphics[width=6.3cm]{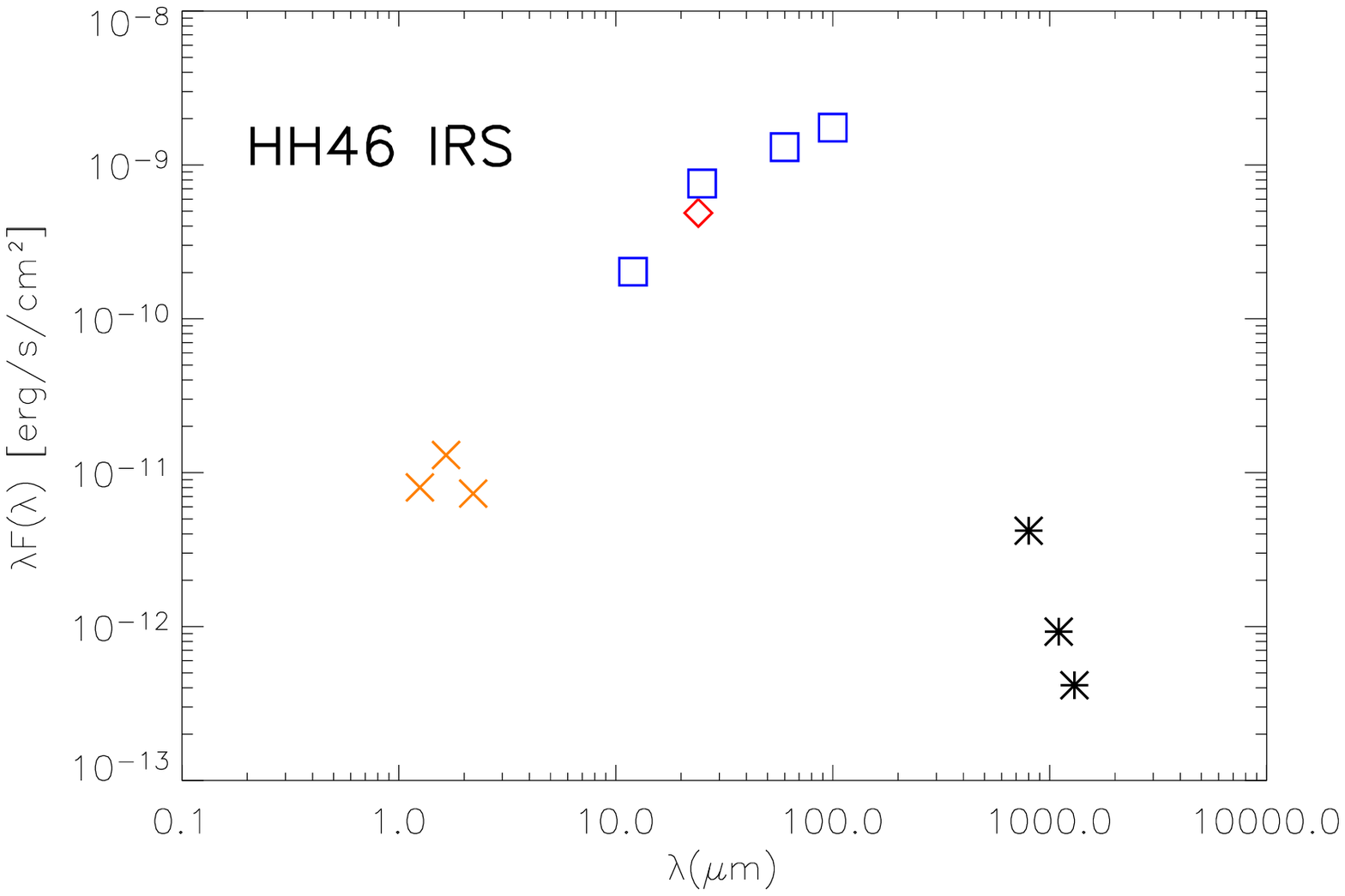}
\vspace{0.2cm}
\end{figure}

\begin{table}[!th]
\caption{\label{t-sed} Photometric measurements available in the literature.}
\begin{center}
\begin{footnotesize}
\begin{tabular}{l l l l}
\hline
\hline
Measurements				&HH26 IRS		&HH34 IRS	& HH46 IRS	\\
\hline
2MASS J	(mag)				&16.77			&15.06			&14.20			\\
2MASS H	(mag)				&14.07			&13.60			&12.88			\\
2MASS K	(mag)				&11.88			&12.38			&12.72			\\
\hline
IRAS 12 $\mu$m (Jy)		&1.93 			&0.65			&0.81			\\
IRAS 25 $\mu$m (Jy)		&5.13 			&8.49			&6.31			\\
IRAS 60 $\mu$m (Jy)		&20.87		 	&31.7			&26.31			\\
IRAS 100 $\mu$m (Jy)		&67.93			&94.8			&58.27			\\
\hline
MSX 8.28 $\mu$m (Jy)	 & \ldots				&0.19			& \ldots				\\
MSX 14.65 $\mu$m (Jy)	& \ldots				&1.54			& \ldots				\\
MSX 21.30 $\mu$m (Jy)	& \ldots				&5.54			& \ldots				\\
\hline
SPITZER 24 $\mu$m (Jy)	&1.8$^{(8)}$	&6.7$^{(8)}$	&3.9$^{(7)}$   \\
SPITZER 70 $\mu$m (Jy)	&7.9$^{(8)}$	&23.3$^{(8)}$	& \ldots				\\
\hline
350 $\mu$m	(Jy)			&6.30$^{(3)}$	& \ldots			& \ldots				\\
800 $\mu$m	(Jy)			& \ldots				& \ldots				&1.12$^{(4)}$	\\
850 $\mu$m	(Jy)			&1.19$^{(2)}$	&3.4$^{(1)}$	& \ldots				\\
1.1 mm (Jy)						&\ldots				&\ldots				&0.34$^{(4)}$	\\
1.3 mm (Jy)						&0.32$^{(3)}$	&0.42$^{(6)}$	&0.18$^{(6)}$  \\
3.6 cm (mJy)					&\ldots				&0.16$^{(5)}$	&\ldots				\\
\hline
\end{tabular}
\end{footnotesize}
\end{center}
\vspace{0.2 cm}
\small{References. (1) \citet{johnstone06b}; (2) \citet{johnstone01}; (3) \citet{lis99}; (4) \citet{correia98};
(5) \citet{rodriguez96}; (6) \citet{reipurth93}; (7) \citet{noriega-crespo04}; (8) Spitzer archive.}\\*
\end{table}


\begin{table}[!h]
\begin{center}
\caption{\label{lineshh26}Emission features in HH26 IRS.}
\vspace{.2cm}
\begin{tiny}
\begin{tabular}{l c c l}
\hline
\hline
$\lambda$ & $F \pm \Delta F$ & $FWHM$  & ID   \\
($\mu$m)	 & ($10^{-16}$\,erg\,s$^{-1}$\,cm$^{-2}$)&(km\,s$^{-1}$)&\\
\hline
1.6116		&	1.2	$\pm$	0.7		&	140	  &Br 13			     \\
1.6413		&	3.6	$\pm$	1.6		&	235	  &Br 12		     \\
1.6530		&	1.0	$\pm$	0.3		&	100	  &?				     \\
1.6756		&	0.7	$\pm$	0.2		&	45	  &?				     \\
1.6812		&	2.2	$\pm$	0.5		&	100	  &Br 11			     \\
1.6877		&	1.9	$\pm$	0.3		&	60	  &H$_2$ 1-0 S(9)	\\
2.0874		&	2.4	$\pm$	0.5		&	70	  &?				     \\
2.1216		&	69.0	$\pm$	0.4		&	50	  &H$_2$ 1-0 S(1)	\\
2.1541		&	6.9	$\pm$	1.2		&	210	  &	  H$_2$ 2-1 S(2)		     \\   
2.1663		&	26.0	$\pm$	0.8		&	140	  &Br $\gamma$  		     \\
2.1791		&	4.6	$\pm$	0.5		&	80	  &\ion{Ti}{i} \\
2.1906		&	3.4	$\pm$	0.6		&	80	  &\ion{Ti}{i} \\
2.1905		&	3.1	$\pm$	0.5		&	70	  &		  ?	  \\
2.2012		&	2.3	$\pm$	0.4		&	60	  &H$_2$ 3-2 S(3)		     \\
2.2065		&	4.5	$\pm$	0.6		&	80	  &\ion{Na}{i} \\
2.2092		&	7.8	$\pm$	0.6		&	95	  &\ion{Na}{i}\\
2.2231		&	15.9	$\pm$	0.4		&	60	  &H$_2$ 1-0 S(0)		     \\
2.2391		&	2.3	$\pm$	0.5		&	65	  &?				     \\
2.2475		&	6.6	$\pm$	0.4		&	75	  &H$_2$ 2-1 S(1)		     \\
2.2940		&	200.2	$\pm$	1.5	&	\ldots 		  &CO (2-0)$^(a)$		  \\
\hline
\end{tabular}
\end{tiny}\\
\end{center}
\small{Notes. Fluxes are not corrected for extinction. (a) measured up to 2.305 $\mu$m.}
\end{table}

\begin{table}[!h]
\begin{center}
\caption{\label{lineshh34}Emission features  in HH34 IRS.}
\vspace{.2cm}
\begin{tiny}
\begin{tabular}{l c c  l}
\hline
\hline
$\lambda$ & $F \pm \Delta F$ & $FWHM$ & ID   \\
($\mu$m)	 & ($10^{-16}$\,erg\,s$^{-1}$\,cm$^{-2}$)&(km\,s$^{-1}$)&\\
\hline
1.5773	&   	    1.5	    $\pm$	0.3   &		85	&    ?  			      \\
1.5888	&   	    4.2	    $\pm$	0.7   &		265	&    Br14			       \\
1.5895	&   	    2.3	    $\pm$	0.3   &		90	&    ?  			       \\
1.5996	&   	    37.6    $\pm$	0.4   &		100	&    [\ion{Fe}{ii}]			       \\
1.6115	&   	    4.3	    $\pm$	0.6   & 	180	&    Br13			       \\
1.6413	&   	    7.9	    $\pm$	0.1   &	        290	&    Br12			       \\
1.6437 	&   	    156.0   $\pm$	1.0   &		100	&   	       [\ion{Fe}{ii}]   \\
2.1220	&   	    22.0    $\pm$	0.1   &		15	&    H$_2$   1-0 S(1)		       \\
2.1329	&   	    8.7     $\pm$	0.5   &		80	&    [\ion{Fe}{ii}] \\
2.1545	&   	    1.1     $\pm$	0.3   &		60	&    H$_2$   2-1 S(2)		       \\
2.1663 	&   	    39.0    $\pm$	0.1   &		240	&   	Br $\gamma$	       \\
2.1793 	&   	    2.5	    $\pm$	0.5   &		55	&    \ion{Ti}{i}			       \\
2.1902 	&   	    3.5     $\pm$	1.1  	&	170	&    \ion{Ti}{i}? \\
2.2016 	&   	    1.2	    $\pm$	0.3   &		45	&    H$_2$ 3-2s(3)		       \\
2.2065 	&   	    8.3	    $\pm$	0.7   &		115	&    \ion{Na}{i} \\
2.2092 	&   	    7.9     $\pm$	0.7   &		110	&    \ion{Na}{i} \\
2.2234	& 	    9.1     $\pm$    	0.2   &	 	10	&    H$_2$  1-0 S(0)		       \\
2.2239 	&   	    34.0    $\pm$	0.7   &		95	&    [\ion{Fe}{ii}] \\
2.2403 	&   	    2.3     $\pm$	0.4   &		55	&    ?  			      \\
2.2438 	&   	    6.3	    $\pm$	0.6   &		85	&    [\ion{Fe}{ii}] \\
2.2478 	&   	    2.2	    $\pm$	0.2   &		10	&    H$_2$  2-1 S(1)		       \\
2.2536 	&   	    8.8     $\pm$	0.5   &		85	&    [\ion{Fe}{ii}] \\
2.2807 	&   	    2.1	    $\pm$	0.6   &	75		&     ? 			       \\
2.2940 	&   	    184.0   $\pm$	1.5   &	\ldots		&    CO (2-0)$^(a)$		       \\
\hline
\end{tabular}
\end{tiny}
\end{center}
\small{Notes. Fluxes are not corrected for extinction. (a) measured up to 2.305 $\mu$m.}
\end{table}

\begin{table}[!h]
\begin{center}
\caption{\label{lineshh46}Emission features  in HH46 IRS.}
\vspace{.2cm}
\begin{tiny}
\begin{tabular}{l c c l}
\hline
\hline
$\lambda$ & $F \pm \Delta F$ & $FWHM$ & ID   \\
($\mu$m)	 & ($10^{-16}$\,erg\,s$^{-1}$\,cm$^{-2}$)&(km\,s$^{-1}$)&\\
\hline
1.5889		&	5.1	$\pm$	0.8			&	310	&  Br 14		  \\
1.5989		&	18.3	$\pm$	0.3			&	60	&  [\ion{Fe}{ii}] \\
1.6114		&	2.7	$\pm$	0.4			&	150	&  Br 13			  \\
1.6387		&	12.1	$\pm$	0.5			&	100	&  ?				  \\
1.6411		&	21.1	$\pm$	0.5			&	120	&  Br 12			  \\
1.6429		&	173.2	$\pm$	0.4			&	65	&  [\ion{Fe}{ii}]	  \\
2.0847		&	1.3	$\pm$	0.4			&	80	&  	  ?			  \\
2.1218		&	14.8	$\pm$	0.3			&	45	&  H$_2$ 1-0 S(1)		  \\
2.1321		&	2.4	$\pm$	0.4			&	110	&  [\ion{Fe}{ii}]			  \\
2.1647		&	3.0	$\pm$	0.3			&	80	&  ?		  \\
2.1662		&	7.4	$\pm$	0.6			&	175	&  Br $\gamma$  		  \\
2.2231		&	20.3 	$\pm$	0.7			&	80	&  H$_2$ 1-0 S(0)	  \\
2.2432		&	5.1	$\pm$	2.0			&	350	&  [\ion{Fe}{ii}]		  \\
2.2477		&	2.5	$\pm$	0.5			&	50	&  H$_2$ 2-1 S(1)	  \\
2.2526		&	3.4	$\pm$	0.9			&	85	&   [\ion{Fe}{ii}]			  \\
2.2936		&	44.8	$\pm$	0.9			&	\ldots	&  CO (2-0)$^(a)$		  \\
\hline
\end{tabular}
\end{tiny}
\end{center}
\small{Notes. Fluxes are not corrected for extinction. (a) measured up to 2.305 $\mu$m.}
\end{table}


\begin{table}[!b]
\caption{\label{t-vel} Radial velocities ($V_\mathrm{LSR}-V_\mathrm{cloud}$) measured on the brightest lines of the detected series.}
\begin{center}
\begin{footnotesize}
\begin{tabular}{l c c c c}
\hline
\hline
Source			&	\ion{H}{i} Br$\gamma$	 		& \ion{Fe}{ii} (1.64$\mu$m)  & H$_2$ (2.12$\mu$m) &  \ion{Na}{i}\\
					&	(km\,s$^{-1}$)					&			(km\,s$^{-1}$)			&		(km\,s$^{-1}$)			&		(km\,s$^{-1}$)			\\
\hline 
HH 26 IRS		&	-3							&	\ldots					&	-56				&	-10			\\
HH 34 IRS    &	-3							&	-88$^{(a)}$				&	-11				&	3				\\
HH 46 IRS    &	-3							&	-200				&	-11				&	\ldots				\\
\hline
\end{tabular}
\end{footnotesize}
\end{center}
\small{Notes. We estimate an uncertainty of 5 km\,s$^{-1}$ for the values here reported.
(a) the line is actually double-peaked, with a blue peak $V_\mathrm{LSR}-V_\mathrm{cloud}$=$-93.5$ km\,s$^{-1}$ and a red peak $V_\mathrm{LSR}-V_\mathrm{cloud}$=$-62.5$ km\,s$^{-1}$.}
\end{table}


\subsection{Bolometric luminosity}

The total luminosity $L_\mathrm{bol}$ of the sources has been derived by integrating the observed spectral 
energy distributions that include measurements spanning from the near IR to the mm region (see table \ref{t-sed}). The calculation is 
performed starting from the $J$ band value and considering straight lines (in the Log($\lambda$)-Log($\lambda F_\lambda$) plan) between available SED points; a final correction at the longest wavelengths is applied assuming that the emission $F_\mathrm{\lambda}$ is decreasing as 1/$\lambda^2$ after the last available observation at 1.3 mm.  
The derived value might be subject to a bias because of the possible presence of unseen companions contributing to the observed flux or contamination from other stars in the field, especially for the IRAS measurements at 60 and 100 microns.
For this reason, we considered recent Spitzer-MIPS images to verify the crowding of the fields at mid IR wavelengths and derive more accurate flux estimates. Results for HH46 IRS at 24 $\mu$m are given by (\citealt{noriega-crespo04}, while for HH26 IRS and HH34 IRS we have used available archive images at 24 and 70 $\mu$m. Spitzer photometry at 24 and 70 microns was considered instead of the respective IRAS fluxes at 25 and 60 microns.
 
For HH26 IRS, Spitzer photometry shows that contamination from nearby sources is indeed very likely in the IRAS measurements. In fact, the calculated 70 $\mu$m flux is much lower than the IRAS 60 $\mu$m point. A luminosity of 9.2 $L_{\sun}$ was derived considering the IRAS 100 $\mu$m point that however is likely to be overestimated. Neglecting the 100 point we obtains 4.6 $L_{\sun}$. These results are to be regarded as an upper and lower limit to the total luminosity of the source. 

HH34 IRS does not appear in the IRAS point source catalogue, probably owing to source confusion problems; however, IRAS fluxes were extracted 
by \citet{cohen87}. Actually, the Spitzer images show that important contamination from other sources is not likely, as the source appears to be quite isolated in the field. On the other hand, the SED displays a very rapid increase at 100$\mu$m with respect 
to the 60$\mu$m point. The derived Spitzer fluxes are lower than the IRAS fluxes, although the discrepancy is not as big as in the case of HH26 IRS.
The derived total luminosity is about 19.9 $L_{\sun}$ considering the IRAS points and 12.4 $L_{\sun}$ neglecting the 100 $\mu$m measurement.

\citet{reipurth00} found HH46 IRS to be a binary using HST/NICMOS observations. The binary flux ratio in the band of observation is about 0.7, 
but it is impossible to determine which of the two components is responsible for the jet. Also in this case, the Spitzer flux at 24 microns \citep{noriega-crespo04} is lower than the one measured by IRAS.
The inferred total luminosity of the binary is 15.0 $L_{\sun}$, which of course is an upper limit to the bolometric luminosity of the HH46 exciting source.

\subsection{Source classification}

The near-IR photometry and the availability of Spitzer data allow us to refine the classification of the three sources based on their SEDs.
The derived $K$-[24] colours are 10.3 mag for HH26 IRS, 12.5 mag for HH34 IRS and 13.3 mag for HH46 IRS. 
These values indicate that the three sources can be indeed considered as Class I or younger, following \citet[][in press]{rebull07} who classify as Class I sources those having a $K$-$Q$ colour larger than 8.3. 
The near-IR $J-H$ and $H-K$ colours of HH34 IRS and HH26 IRS are instead  typical of embedded T Tauri stars \citep{meyer97}.
The HH46 IRS colours are rather peculiar, since the source falls on the left of the reddened main-sequence star locus. As we mention in Sect. \ref{par-extinct}, however, the displacement of the sources in this colour-colour diagram is not indicative of the true amount of extinction, since the presence of a large quantity of scattered light has the effect of making the star appear bluer and less reddened \citep[see e.g][]{massi99, stark06}. 
Contamination due to light scattered by the circumstellar disc is indeed very likely in all our sources, that have estimated disc inclination angles, with respect to the line of sight, of 20-30$^\circ$.  

Indication that the sources are in an early stage of evolution is also given by their relatively massive dust envelopes \citep[0.2--1 M$_{\sun}$][]{reipurth93,lis99}. 
Indeed, the ratio between  $L_\mathrm{bol}$ and the sub-mm luminosity measured longward of 350\,$\mu$m ($L_\mathrm{submm}$), estimated from the photometric data, is rather small in all the sources ( $\sim$60 for HH26 IRS, $\sim$ 70 for HH34 IRS and $\sim$160 for HH46 IRS), which would indicate that they actually satisfy the Class 0 criterium  $L_\mathrm{bol}/L_\mathrm{submm}<200$ as defined by \citet{andre93}. 
Moreover, comparing the observed  SEDs with the grid of YSO SED models by \citet{whitney03} and \citet{robitaille06}, 
we derive that models for evolved Class 0 sources provide better fits to the observational data than those for Class I/IIs, 
although some parameters,  such as the depth of the silicate feature, largely deviate with respect to the observed values.
In any case, these findings strongly suggest  that the sources are indeed very young and are not more evolved (Class II) objects.

\section{Analysis and physical parameters}

\subsection{Extinction, Accretion and stellar parameters}
\label{par-extinct}

The fraction of the total luminosity due to accretion ($L_{\mathrm{acc}}$) must be determined in order to get an estimate of the mass accretion rate.
\citet{nisini05a} derived this quantity for a sample of young sources directly determining the spectral type and veiling from the analysis of the absorption features from the stellar photosphere, then inferring the stellar parameters and luminosity $L_{\mathrm{*}}$, from which the accretion luminosity can be obtained ($L_{\mathrm{acc}} = L_{\mathrm{bol}} - L_{\mathrm{*}}$).
However, our spectra show no evidence of well detected ($S$/$N$\,$\gtrsim$\,3) absorption features from the photosphere,
indicating the presence of a strong veiling, so that an alternative method has to be used to derive the accretion luminosity.

We have therefore considered the relationship found by \citet{muzerolle98} on the basis of observations of T Tauri stars, in which $L_{\mathrm{acc}}$ is directly related to the (extinction-corrected) \ion{H}{i} Br$\gamma$ flux:
\begin{equation}
\mathrm{Log} \, \frac{L_{acc}}{L_{\sun}} = (1.26) \, \mathrm{Log} \, \frac{L_{Br\gamma}}{L_{\sun}}+(4.43)
\end{equation}
Such an approach implies considering this relation still valid for the more embedded sources of our sample; the only observational evidence for such an assumption has been given so far for three sources in the R CrA star-forming region \citep{nisini05a}.

In order to retrieve absolute Br$\gamma$  luminosities from observed fluxes, it is necessary to correct them for the extinction and thus to measure the value of $A_\mathrm{K}$ toward the sources. 

Estimates of the extinction can be inferred adopting standard methods based on the analysis of the spectral features. 
For HH46 IRS, considering the silicate absorption feauture at 9.7 $\mu$m observed by Spitzer \citep{boogert04} and using the relation by  \citet{mathis98}, we obtain an $A_{V}$ value between 32 and 38 mag, corresponding to an $A_{K}$ 3.5-4.2 adopting the \citet{rieke85} extinction law.
In the case of HH26 IRS, measurements of the ice absorption feature at 3 $\mu$m \citep{simon04} and the relationship by \citet{tanaka90} provide an $A_{V}$ value of the order of 29 mag ($A_{K} \sim 3.2$).

We note, however, that the ISM relationships between dust features and extinction, derived for the ISM, may not be strictly valid for dust in the circumstellar environment of YSOs \citep[e.g.][]{chiar07}.

Finally, for HH34 IRS 
only a lower limit of $A_\mathrm{V}$=7 mag can be taken adopting the value recently estimated by \citet{podio06} from the ratio of [\ion{Fe}{ii}] lines excited in the jet region at a distance less than $\sim$1\arcsec from the source.

Given these problems, since a reliable determination of the extinction is crucial in order to derive the accretion luminosity from Br$\gamma$, we have tried to use a different and independent method to compute ``self-consistent'' $A_K$, based on the observed magnitudes and on reasonable assumptions on the stellar properties of the objects.
Indeed, the intrinsic stellar luminosity is given by:
\begin{equation}
Log L_*/L_{\sun}=-0.4(M_\mathrm{bol}-M_\mathrm{bol,\sun})
\end{equation}
where
\begin{equation}
\label{mbol}
M_\mathrm{bol}=BC+M_K+(V-K)_*
\end{equation}
The absolute magnitude $M_K$ in Eq. \ref{mbol} depends on the observed magnitude $m_K$, the extinction $A_K$, the veiling $r_K$ and the distance modulus DM:
\begin{equation}
M_K=m_K+2.5Log(1+r_K)-A_K-DM
\end{equation}
while the values of the bolometric correction $BC$ and the intrinsic stellar colour $(V-K)_*$ depend on the spectral type and age of the objects. Thus,if we assume that our sources are late-type stars located on the birthline and use the parameters provided by a set of evolutionary models, we can obtain an estimate of $A_K$ from the observed $K$-band magnitude and veiling of the objects. 

On this basis, we have adopted a procedure allowing us to derive the different physical quantities in a self-consistent way.
Namely, we have varied the $A_K$ until we have found a value able to 
reconcile the stellar luminosity given by Eq. 2 with the value given by $L_{\mathrm{bol}}-L_{\mathrm{acc}}$, where 
$L_{\mathrm{acc}}$ is derived from the extinction-corrected (by the considered $A_K$) $Br\gamma$ line emission using Eq. 1.

For example, 
we can easily derive that an extinction value $A_V$ = 29 mag for HH26 IRS (such as the one derived from the ice feature) would lead to a stellar luminosity which is smaller than the value given by L$_{\mathrm{bol}}-$ L$_{\mathrm{acc}}$ ($\Delta$ L$_\mathrm{*} \sim 2.1 $L$_{\sun}$), pointing to an inconsistency in the independently inferred $A_K$ and $L_{\mathrm{*}}$ values.

We have carried out the described procedure assuming the birthline location of \citet{palla93} and considering the colours and bolometric correction given by \citet{siess00} to convert $L_{\mathrm{*}}$ into $M_K$ (Eq. 2 and 3);
thus, we have found the extinction providing ``consistent'' $L_{\mathrm{*}}$ values.

 We derive for HH26 IRS an $A_{K}=4.1$ mag with a spectral type K7 ($L_{\mathrm{*}}=3.7$ L$_{\sun}$, $R_{\mathrm{*}}$=4.1 R$_{\sun}$,  $M_{\mathrm{*}}$=0.6 M$_{\sun}$) and for HH34 IRS an $A_{K}=4.9$ mag  with a spectral type M0 ($L_{\mathrm{*}}=2.9$ L$_{\sun}$, $R_{\mathrm{*}}$=3.9 R$_{\sun}$,  $M_{\mathrm{*}}$=0.5 M$_{\sun}$). 

In the case of HH46 IRS we know that the source is a binary, but the contribution from the companion remains unknown.
Assuming that the bolometric luminosity is equally distributed between the two sources, the best results are obtained considering a K5 star with  $A_{K}=4.9$ mag ($L_{\mathrm{*}}$=6.0  L$_{\sun}$, $R_{\mathrm{*}}=$4.2 R$_{\sun}$,  $M_{\mathrm{*}}$=1.2 M$_{\sun}$). 
If we assume that the contribution to the bolometric luminosity of our considered target is larger with respect to the companion, we find that it is not possible to obtain consistent results for $L_{\mathrm{*}}$ smaller than 10 L$_{\sun}$; exploring the possibility of a more massive star, we can still find consistency with an $A_{K}\sim$5.3 mag and a spectral type G9/K0 ($L_{\mathrm{*}}=12.5$ L$_{\sun}$, $R_{\mathrm{*}}$=4.3 R$_{\sun}$,  $M_{\mathrm{*}}$=2.8 M$_{\sun}$). However in this case the bolometric contribution of the companion (considering a derived $L_{\mathrm{acc}}\sim 2.4$ L$_{\sun}$) would be basically zero, which appears unlikely.

In Tab. \ref{param1} we report the parameters derived for the three sources adopting the analysis just described.

Providing self-consistent values of the parameters, we are confident that these are better determinations of the extinction with respect to the previous ones, also taking into account the already mentioned problems that may arise using the relationships established for the ISM for dust around YSOs.

Moreover, adopting the extinction values from spectral features and jet line ratios, we would infer accretion luminosities $\sim$1 L$_{\sun}$ for HH26 IRS, $\sim$0.1 L$_{\sun}$ for HH34 IRS and 0.3-0.7 L$_{\sun}$ for HH46 IRS, that  represent only a small fraction of the total luminosity of the sources ($\sim$1--20\%), thus implying that most of the source luminosity should be contributed by the stellar photosphere. However, this is in contrast with the early evolutionary state of the objects and with the large $K$ band veiling values we find.

Conversely, the accretion luminosities that we infer adopting our ``self-consistent'' extinction values are 3.2 L$_{\sun}$ for HH26 IRS, 13.3 L$_{\sun}$ for HH34 IRS and 1.5 L$_{\sun}$ for HH46 IRS. The estimated accretion to total luminosity ratios are therefore $\sim$0.5 for HH26 IRS, $\sim$0.8 for HH34 IRS, and $\sim$0.2 for HH46 IRS  (see Tab.\ref{param1}), where we have considered a mean value of the $L_{\mathrm{bol}}$ ranges given in Tab. \ref{targets} for HH26 IRS and HH34 IRS.

From these accretion luminosities we have then derived an estimate of the mass accretion rate $\dot{M}_{\mathrm{acc}}$, using the formula for disc accretion \citep{gullbring98}:
\begin{equation}
\dot{M}_\mathrm{acc}= \frac{L_{\mathrm{acc}} R_{\mathrm{*}}}{GM_{\mathrm{*}}}\, \left(1-\frac{R_{\mathrm{*}}}
{R_{\mathrm{i}}}\right)^{-1}
\end{equation}
where $R_{\mathrm{*}}$ and $M_{\mathrm{*}}$ are the stellar radius and mass, respectively, and $R_{\mathrm{i}}$ is the inner
radius of the accretion disc.
We consider the masses and radii derived from Siess models during the computation of the extinction described above 
and assume a value of $R_{\mathrm{i}}$ = 5 R$_{\mathrm{*}}$ for the disc inner radius \citep[e.g.][]{gullbring98}.
The inferred mass accretion rates are listed in Tab. \ref{param1} and are of the order of $10^{-7}$ (HH26 IRS and HH46 IRS) and 
$10^{-6}$ (HH34 IRS) M$_{\sun}$\,yr$^{-1}$. 

All previous calculations were carried out assuming the measured lower limits on veiling. Considering higher $r_K$ values and reperforming the computations leads to increased $A_K$ and $L_{\mathrm{acc}}$, lower $L_{\mathrm{*}}$ and $M_{\mathrm{*}}$, and consequently to greater $L_{\mathrm{acc}}/L_{\mathrm{bol}}$ and $\dot{M}_\mathrm{acc}$. 
In particular, in the case of HH26 IRS, taking into account a veiling $r_K=4$ we get a higher $A_K\sim4.4$ with a spectral type M1 and $L_{\mathrm{acc}} \sim 4.1 L_{\sun}$; we would thus obtain $L_{\mathrm{acc}}/L_{\mathrm{bol}} \sim 0.6$ and $\dot{M}_\mathrm{acc} \sim 1.6 \cdot 10^{-6}$ M$_{\sun}$\,yr$^{-1}$.

\subsection{Jet mass flux}
\label{jmf}
The jet mass flux can be derived from emission line measurements, providing that the jet is resolved and 
its velocity is known \citep[see e.g.][]{hartigan95,nisini05b}. For our purpose of directly 
comparing the mass loss with the mass accretion rate, we need to 
determine the mass flux relative to jet component as close as possible to the central source. Jet knots far 
from the protostars may indeed be related to older episodes of mass ejection, which could be associated to a different mass flux.

For HH46, we have estimated the mass flux from the luminosity of the [FeII] 1.64 $\mu$m line,
 measured on the first 3 arcsec of the jet length. There are no measurements of the jet diameter in this inner jet section,
which is not optically visible: at a distance of $\sim$5 arcsec, a jet diameter of $\sim$ 1 arcsec has been
measured by \citet{eisloeffel94b}. Since jet diameters increase with distance from the driving source 
\citep[e.g.][]{dougados00}, the diameter in the portion of the jet we are considering should be $\sim$ 0.6 arcsec. 
Thus, in addition to the flux losses due to the seeing, we have also taken into account a correction factor of 2 to retrieve an estimate of the 
total flux in the considered jet section.
We therefore have:

\begin{equation}
\label{eq-mloss}
\dot{M}_\mathrm{loss} = \mu\,m_H\times(n_H\,V)\times v_{t}/l_{t}
\end{equation}
and
\begin{equation}
n_H\,V = L_{1.64\mu m}\,\left(h\,\nu\,A_{i}\,f_{i}\,\frac{[Fe]}{[H]}\right)^{-1}
\end{equation}
where  $A_{i}$, $f_{i}$ are the radiative rate and fractional population of the upper level of the considered transition, 
$[Fe]/[H]$ is the total Fe abundance with respect to hydrogen (assumed 2.82 $\cdot 10^{-5}$, Asplund 2005\nocite{asplund05}) and
$v_{t}$ and $l_{t}$ are the velocity and length of the knot, 
projected perpendicularly to the line of sight. We assume here that all Fe is ionized, 
as it is expected in the excitation conditions of HH objects \citep{nisini02}. 
The fractional population has been computed adopting a NLTE statistical equilibrium code 
\citep{nisini02} with an electron density $n_e$=3.7\,10$^{3}$ cm$^{-3}$, estimated from the 
[\ion{Fe}{ii}]1.64/1.60 observed line ratio. The electron temperature has been assumed 10$^4$ K. The  tangential velocity
has been inferred through the proper motion study of \citet{eisloeffel94b}, and it 
is equal to 170 km\,s$^{-1}$. The 1.64 $\mu$m line luminosity has been calculated from the observed 
flux, which has been dereddened assuming that the extinction in the considered jet knot is
A$_{V}$ = 6.6 mag. This value has been estimated from the [\ion{Fe}{ii}] 1.64/1.25 $\mu$m line ratio measured 
by \citet{fernandes00}, adopting the expected theoretical ratio of \citet{quinet96}.
The resulting mass flux value is about 3$\cdot$10$^{-8}$ M$_{\sun}\,$yr$^{-1}$.
Such a value is about a factor of 3 and 10 lower than the values estimated by \citet{bacciotti99} and by \citet{hartigan94}, respectively,  through the measurement of the total
gas density, and thus the total mass in the jet, following the procedure known as the BE technique 
(from Bacciotti \& Eisl\"offel, 1999),  or adopting a shock model. It has been shown that the  mass flux thus inferred may be overestimated due to the assumption that  the considered jet area is filled with gas at the derived density \citep[][]{hartigan94,nisini05a}. 
If the shock  is caused by jet time-variability, the lack of correction for shock compression will overestimate the time-averaged $\dot{M}_\mathrm{loss}$ by a factor of about 5 \citep{hartigan94}.
On the other hand, 
\citet{hartigan94} also compute a mass flux value of 4.2$\cdot$\,$10^{-7}$ M$_{\sun}\,$yr$^{-1}$ 
from the luminosity of the [\ion{O}{i}] 6300\AA\ line, adopting a relationship similar to that given in Eq. \ref{eq-mloss}, although on a jet region much more distant ($\sim$10-20\arcsec from the source) than the one we consider.

Part of the above discrepancies may arise from our assumption of solar abundance of gas-phase Fe. This
assumption may not be valid if part of iron
is still locked in grains along the jet. This is indeed what has been found in other jets (HH1 and HH34) by \citet{nisini05b} 
and \citet{podio06}, who showed that in the inner jet regions only a fraction between 30 and 70\%
of all iron may be in gaseous form. If the same applies also to the HH46 jet, than our data give a mass flux up to
$\sim$2$\cdot10^{-7}$ M$_{\sun}\,$yr$^{-1}$.

The mass flux in the HH34 inner jet has been measured by several authors.  The most recent determinations were given by 
Davis et al. (2003), who used the luminosity of the [\ion{Fe}{ii}]1.64$\mu$m line, by \citet{podio06}, who compared 
different techniques employing both optical and IR lines, and finally by \citet{garcia_lopez07}, who used the
[\ion{Fe}{ii}]1.64$\mu$m luminosity from the same set of data we are here analysing.
All these determinations agree on a value in the range 4-7$\cdot$10$^{-8}$ M$_{\sun}$\,yr$^{-1}$. If we correct
the determinations based on [\ion{Fe}{ii}] line luminosity by the 30\% of iron dust depletion found in the inner jet knot
by  \citet{podio06}, we get a mass flux up to $\sim$1.2\,10$^{-7}$ M$_{\sun}$\,yr$^{-1}$, which is in agreement
also with the determinations by Hartigan et al.(1994).

The jet in HH26 is seen close to the star only in H$_2$, indicating that the gas is
mainly molecular and that the atomic component does not give a significant contribution
to the transport of material outwards. We have therefore measured the mass flux from the 
luminosity of the observed 2.12$\mu$m H$_2$ line. More explicitly, Eq. 6 can be here written as:
\begin{equation}
\dot{M}_\mathrm{loss} = 2\mu\,m_H\times(N(H_2)\,A)\times v_{t}/l_{t}
\end{equation}
where $N(H_2)$ is the column density of molecular hydrogen and $A$ is the considered emission area. 
Since H$_2$ is optically thin and quickly thermalised, the total column density can be directly measured from the
dereddened line luminosity:
\begin{equation}
N(H_2)\,A = L_{2.12\mu m}\,(h\,\nu\,A_{i}\,f_{i})^{-1}
\end{equation}
with $f_{i}$ given by the LTE Boltzman population at the gas kinetic temperature.
A temperature of $\sim$ 2000 K can be estimated by the ratio of the various observed lines. 
We have considered only the first knot of the HH26 microjet, which extends within about 2 arcsec from the 
central source.
As in the case of HH46, we have applied a correction of a factor of 3 to take into account the width of the jet at the knot position, 
which is of $\sim$1\arcsec \citep{chrysostomou07}. The adopted line flux results consistent with the one measured by \citet{giannini04} with a slit of 1\arcsec .
The tangential velocity has been estimated to be around 130 km\,s$^{-1}$ 
from the measured radial velocity of -60 km\,s$^{-1}$, assuming an inclination angle of 65$^\circ$ with respect to the line of sight 
\citep{davis97}.
Here, a critical parameter is the reddening towards the jet. \citet{giannini04} 
measured an $A_\mathrm{V}$ =2  on the external HH26A knot; it is however likely that the extinction has a sharp 
increase close to the object, since a value of $A_V \sim$38 has been
estimated on-source. A method to have an estimate of $A_V$ in the H$_2$ emitting region is to 
construct a Boltzman diagram with  the H$_2$ lines detected on source (Table 3) and take the $A_V$ value that gives the better alignment of the different transitions on a straight line (see e.g. Nisini et al. 2002). 
Given that the considered lines do not cover a wide range of wavelengths, the method is not very sensitive to small $A_V$ variations: we find that the best alignment is obtained with $A_V$ values between 10 and 20 mag. With this range of extinctions, we get  H$_2$ column densities ranging between 
5 and 13$\cdot$10$^{18}$ cm$^{-2}$, corresponding to a mass flux ranging
from $\sim$2$\cdot$10$^{-8}$ to 5.3$\cdot$10$^{-8}$ M$_{\sun}$\,yr$^{-1}$. 

The (one-sided) estimates of $\dot{M}_\mathrm{loss}$ we have thus derived are summarised in Tab. \ref{param2}.

\begin{table*}
\begin{center}
\caption[]{Derived parameters for the observed targets.}
\begin{small}
\begin{tabular}{l c c c c c c c c c}
\hline 
\hline
Source         &  $A_K$	&$L_\mathrm{acc}$&$L_\mathrm{*}$& $M_\mathrm{*}$ &ST &$L_\mathrm{acc}/L_\mathrm{bol}$	& $\dot{M}_\mathrm{acc}$	&
$\dot{M}_\mathrm{loss}$					& $\dot{M}_\mathrm{loss}/\dot{M}_\mathrm{acc}$	\\
				 	&	(mag)		&(L$_{\sun})$			&(L$_{\sun}$)		& 	(M$_{\sun}$) &	&													    &	(10$^{-7}$ M$_{\sun}$\,yr$^{-1}$)	& (10$^{-7}$ M$_{\sun}$\,yr$^{-1}$)	&																					\\
\hline
HH26 IRS		&	$\sim$4.1     & 3.2   & 	3.7	&0.6& K7 & $\sim$0.5		& 8.5  			& 0.2-0.5	 & 	0.02-0.06 	\\ 
HH34 IRS		&	$\sim$4.9    & 13.3 	& 	2.9	&0.5& M0 & $\sim$0.8		& 41.1 			& 0.4-1.2$^{(a)}$	& 	0.01-0.03		\\ 
HH46 IRS		&	$\sim$4.9	   & 1.5    & 	6.0 	&1.2& K5 & $\sim$0.2		& 2.2			& 0.3-2.0$^{(b)}$	& 0.14-0.90		\\ 
\hline
HH100-IR		&	3.3$\pm$0.3 	&12$\pm$2		&3.1$\pm$0.9	&0.3-0.75& K5-M0 	& 0.80		& 10-20 		& \ldots				&\ldots				\\
IRS2				&	2.4$\pm$0.3 	&7.7$\pm$2.5	&4.3$\pm$1.5 &1.1-1.8 &  K2 		& 0.60	 	& 2-3 			& \ldots				&\ldots				\\
IRS5a			&	5.0$\pm$0.6 	&$\sim$0.4		&1.6$\pm$0.5	&0.4-0.9& 	K5-K7	 & 0.15		& 0.2-0.3 		& \ldots				&\ldots				\\
\hline
\end{tabular}
\end{small}
\label{param1}
\end{center}
\small{Notes. Columns show in order: $K$ band extinction, accretion luminosity, stellar luminosity (assuming the star on the birthline), estimated stellar mass and spectral type (using \citet{siess00} models), accretion to bolometric luminosity ratio, mass accretion rate, (one-sided) mass loss rate, mass loss rate to mass accretion rate ratio. The stellar luminosity and the fraction of the luminosity due to accretion are computed considering the mean value of $L_\mathrm{bol}$ given in Tab. \ref{targets}. The same results for three Class I sources of the R CrA region \citep{nisini05a} are reported for comparison (see text for details). (a) Lower value from optical line determinations \citep{podio06}; upper value from [\ion{Fe}{ii}] 1.64 $\mu$m line luminosity, assuming Fe gas abundance from \citet{podio06}. (b) Upper value determined assuming that only $\sim$30\% of iron is in gaseous form.}
\end{table*}

\begin{table*}
\begin{center}
\caption[]{Derived parameters for the observed targets.}
\begin{small}
\begin{tabular}{l c c c c c c}

\hline 
\hline
Source         &  $r_K$ 				& Br$\gamma$ EW &  Br$\gamma$/NaI 	& Br$\gamma$/H$_\mathrm{2}$	& Br$\gamma$/\ion{Fe}{ii} & CO(2-0)/\ion{Na}{i}	\\
				 	&	      					& ($\AA$)				  &								&													&	 &\\
\hline
HH26 IRS		& $>2$ 				& 		-5.0					& 1.8							& 0.3					& $>$25 & 15	\\
HH34 IRS		& $>5$				    & 		-8.9					&  2.4						& 1.7					&	0.22 & 11.6\\
HH46 IRS		& $>1$				& 		-4.8					& $>$5.1					& 5.0					&	0.05& $>$ 30\\
\hline
HH100-IR		& $6.0\pm0.5$		& 		-7.9					 & 6.9							& 93.0					& \ldots& 4.6\\
IRS2				& $2.9\pm0.2$		& 		-7.0					 & 19.2							& 140.0				& \ldots& 3.5\\
IRS5a			& $1.0\pm0.1$		& 		-0.6					 & \ldots												& \ldots						& \ldots &\ldots\\
\hline

\end{tabular}
\end{small}
\label{param2}
\end{center}
\small{Notes. Columns show in order: $K$ band veiling, Br$\gamma$ equivalent width, Br$\gamma$/\ion{Na}{i}, Br$\gamma$/H$_\mathrm{2}$, Br$\gamma$/\ion{Fe}{ii} and CO(2-0)/\ion{Na}{i} flux ratios. The same results for three Class I sources of the R CrA region \citep{nisini05a} are reported for comparison (see text for details).}
\end{table*}


\section{Discussion}

It is interesting to analyse the derived accretion and ejection properties and their mutual connection in the light of the results provided by some works that have recently investigated the characteristics of embedded protostellar sources.
In particular, we will compare the results here presented to observations carried out by \citet{nisini05a} (hereafter N05) on a small sample of Class I sources in the R CrA star-forming region, using the same instrument and observational setup. 
The most interesting sources (namely HH100-IR and IRS 2) displayed strong \ion{H}{i} emission lines along with well detected absorption features from which stellar properties and veiling could be derived.
Unlike the objects analysed in this paper, all the CrA sources showed no signature of jet emission in the region closely surrounding the central star; only HH100-IR has been suggested to be the driving source of the Herbig Haro objects HH101 and HH99 \citep{hartigan87} and of a bipolar molecular outflow \citep{anglada89}, which however are observed far from the central object.

To help the comparison, we added in Tab. \ref{param1},\ref{param2} also the parameters of the three CrA sources found by N05 to have \ion{H}{i} emission.

We will also refer to two recent surveys of embedded sources that have investigated the nature of Class I sources with respect to the Class II objects, in order to verify the paradigm of the star formation according to which Class I are actively accreting objects less evolved than Class II:
\citet{white04} (hereafter W04) analysed high resolution optical spectra of a large sample of \textit{environmentally young stars} (i.e. Class I and Class II objects powering Herbig-Haro flows) in Taurus-Auriga. However, since the Class Is observed by the authors are optically visible, their sample is probably biassed toward more evolved objects than standard Class I sources.
\citet{doppmann05} (hereafter D05) studied high resolution near IR spectra of a large set of Class I and flat-spectrum sources in various nearby star-forming clouds and derived that Class I show in average higher veilings and angular velocities than what is observed in Class II sources, as expected for less evolved and accreting objects.

\subsection{Spectral features}

The medium resolution spectra of our sources show important different characteristics.
In fact, both the detected emission features and their relative strength vary depending on the objects, as can be seen in Fig. \ref{plots26}, \ref{plots34} 
and \ref{plots46}.
 \ion{H}{i}, CO, and H$_2$ emission is detected in each of three sources, which in general should be suggestive of the presence of accretion/ejection 
flows and of an accretion disc.
These features are usually, but not always, observed in embedded sources; in particular, D05 detected  \ion{H}{i}, CO and H$_2$ emission in 65\%, 15\% 
and 44\% of the objects of their sample, respectively, remarking as the presence of the CO is always accompanied by  \ion{H}{i} emission. This is a 
behaviour we find in our sources as well, as in those of N05.
In contrast, the presence of permitted lines does not appear to be strictly connected with the simultaneous presence of jet lines from the region closely 
surrounding the central source. For instance, the \ion{Na}{i} is detected both in HH26 IRS and HH34 IRS, but the [\ion{Fe}{ii}] lines are visible only in 
the second source.
This lack of correlation is extreme in the case of the CrA objects in N05, with strong permitted emission lines and no jet signatures at all. Therefore, 
ejection and accretion signatures that are detected in the spectra do not seem to be strictly related.

All these findings must denote different physical conditions in the close surroundings of the sources: such an indication is also given by the 
analysis of the flux ratio between Br$\gamma$ and \ion{Na}{i}, which is not constant in the objects (see Tab. \ref{param2}). 
We note that this ratio is smaller in the jet sources with respect to the CrA objects; since H and Na have very different ionisation potentials (13.6 and 5.1 eV, respectively), a variation of the flux ratio must reflect a different ionisation degree in the region traced by the Br$\gamma$ and \ion{Na}{i}.

\ion{Na}{i} lines are in most cases observed in Class I sources where CO emission is detected: 5 objects out of 8 in D05 and 2 out 
of 2 in N05. On the contrary, in the majority of T Tauri stars of D05, both CO and \ion{Na}{i} are observed in absorption. 
Noticeably, in our sample the CO(2-0)/\ion{Na}{i} ratio (Tab. \ref{param2}) follows a trend opposite to the one of Br$\gamma$/\ion{Na}{i} ratio, i.e. it is higher in the jet sources.
CO overtone emission is usually interpreted as coming from the inner gaseous disc, heated at temperatures of 2000 K or more by viscous accretion or magnetohydrodynamic waves \citep[e.g][]{carr93}. \ion{Na}{i} could then be originated in an even hotter and  internal disc region at low ionisation. We note that in both HH26 IRS and HH34 IRS, the FWHM velocities of the \ion{Na}{i} lines are smaller than those of Br$\gamma$ of a  factor of about two. This is an evidence that Br$\gamma$ is associated to a gas component at higher keplerian velocity, e.g. the accretion flow and/or ionised stellar/disc wind.
All such findings may suggest that sources with energetic jets possess massive discs characterised by large columns of warm non-ionised gas.

Finally, the absence of \ion{Fe}{ii} emission in HH26 IR points to the presence of a colder molecular jet  in this source, as also confirmed by the lower Br$\gamma$/H$_2$ ratio observed, which indicates a higher H$_2$ content.

\subsection{Veiling}

The lower limits we set on the $K$ band veiling (see Tab. \ref{param2}) show that at least two sources have veilings greater than the ones typically observed in the more evolved T Tauri stars, for which $r_\mathrm{K} \sim 1-2$ \citep[e.g.][]{folha99,johns-krull01}, in agreement with the values measured on other Class I objects \citep[e.g][]{doppmann05,nisini05a,greene96a}.
Large ($r_\mathrm{K} > 2$) values of the $K$-band veiling are indicative of the presence of inner circumstellar dusty envelopes \citep{greene96a} and have been suggested to be a good indicator of high accretion activity in the source, since excess emission should be generated by the infalling matter \citep[e.g.][]{calvet97}.
D05 have measured $K$-band veilings for Class I and flat-spectrum sources in Ophiuchus that are significantly higher than those of Class II in the same region and assume this as a sign of Class I stars being actually embedded objects undergoing significant mass accretion. Finally,  a positive correlation between amount of veiling and accretion luminosity has been found by N05 in the small sample of Class Is in R Cra.

If the $K$ band veiling is indeed  due to the accretion excess above the stellar photosphere, then one should
expect a correlation between the equivalent widths of accretion tracer lines and the veiling.
This kind of correlations have been indeed found for H$\alpha$ and \ion{Ca}{ii} lines in T Tauri stars \citep[e.g][]{muzerolle98b}.
It is interesting to see if such correlation exists also for Br$\gamma$ in the limited sample of our jet sources and CrA sources from N05.

On the assumption that the ratio between the Br$\gamma$ flux and the $K$ band accretion flux is roughly constant, the equivalent width should indeed only depend on the veiling:
\begin{equation}
EW \approx \frac{F_{Br\gamma}}{F_{K,acc}}\frac{r_K}{1+r_K}
\end{equation}
where it has been supposed that all the $K$ band excess emission is due to accretion ($ F_{K,exc} \approx F_{K,acc}$). From this relationship one should expect the line EW to increase roughly linearly for small $r_K$ values, reaching an asymptotic value equal to $F_{Br\gamma}/F_{K,acc}$ for large values of $r_K$.

\begin{figure}[!h]
\caption{\label{f-veil}Br$\gamma$ equivalent widths is plotted against the $K$ band veiling for the three sources and the CrA sources of \citet{nisini05a}.}
\vspace{.1cm}
\centering
\includegraphics[width=6.3cm]{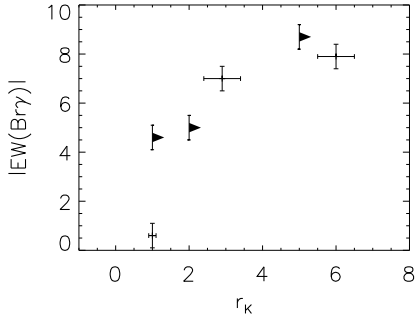}
\vspace{0.2cm}
\end{figure}

Despite the limited number of determinations, and the lower limits on $r_K$ for the jet sources, such a behaviour
is approximately recognisable in Fig. \ref{f-veil}, where  the results for the six considered sources are plotted.
In particular, the large $Br\gamma$ EW measured on HH34 IRS should be noticed, which is similar to the value measured on HH100 IR and in agreement with what is expected given the large lower limit on $r_K$. This indicates that the accretion luminosity in this object should indeed be high, providing support to our estimate of an $A_K$ value of the order of 5 mag.

\subsection{Accretion luminosity}

The sources here analysed have accretion luminosities that are a fraction between  $\sim$0.2 and 0.8 of the total luminosity. In addition to HH34 IRS and HH26 IRS, only three Class I have been found to have a $L_{\mathrm{acc}}/L_{\mathrm{bol}}$ ratio $\gtrsim$ 0.5 so far: HH100-IR and IRS2 in CrA (N05) and YLW15 in Oph \citep{greene02}.
We note that the fraction of $L_{\mathrm{acc}}$ over $L_{\mathrm{bol}}$ in HH46 IRS is somehow uncertain, due to the binary nature of the object.

Recent findings have actually shown that there seem to exist a large spread in the measured accretion luminosities of embedded sources.
Accretion luminosities spanning from a few percent to about 50\% of the total have been measured for instance by W04 in the Taurus-Auriga sources of their sample, although these results (being based on optically visible sources) would better apply to objects in general more evolved than our embedded targets. A similar result has been however recently found also by \citet{beck07} on a sample of Class I in Taurus. Finally, in N05 only two out of  5 investigated Class Is have bolometric luminosities dominated by accretion.  D05 noted that the accretion luminosity derived from Br{$\gamma$} fluxes using Muzerolle's relation for the sources of their sample is low ($10^{-1}<L_\mathrm{acc}/L_{\sun}<10^0$) even in objects with significantly high bolometric luminosities, which could therefore be explained only assuming a very high photospheric contribution. This last result could be partially 
due to an underestimate of the extinction correction to be applied to the Br{$\gamma$}, being the $A_V$ value estimated from near-IR colours. There is nevertheless growing evidence that only a limited number of sources normally classified as Class I actually have accretion-dominated luminosity.

On the basis of the small sample we have gathered, we can try to tentatively define some properties characterising these \textit{Accretion-Dominated Young Objects} (ADYOs), i.e. those sources with $L_{\mathrm{acc}}/L_{\mathrm{bol}}\gtrsim$ 0.5. 
First of all, we note that all the identified accretion-dominated objects have $r_K$ strictly larger than 2. In the D05 sample, only 6 out of 40 sources have such a large veiling, testifying that ADYOs might be just a fraction of all the Class Is.
Sources with massive and dense jets detected very close to the exciting object have high $L_{\mathrm{acc}}/L_{\mathrm{bol}}$ ratios, as we infer from the sample investigated in this paper, but the presence of a jet is not a necessary characteristic of an ADYO (e.g. CrA IRS 2). On the other hand, the fact that the jet sources here investigated have SEDs typical of objects in transitions from Class 0 to Class I, indicates that large accretion luminosities are associated to less evolved sources, while a consistent number of ``classical'' Class I objects may be in a phase of less intense accretion and have already acquired most of their final mass.

We also note that all the ADYOs so far discovered, except for YLW15, have CO and \ion{Na}{i} lines in emission, although there is not a direct correlation between the strength of these features and the accretion luminosity. The majority of the Class Is investigated in D05 as well as in \citet{beck07}, show on the contrary  CO and \ion{Na}{i} features in absorption, which is a further indication that their samples may be biased towards sources lacking massive accretion-heated  inner discs.

\subsection{Mass accretion and loss rates}

The actual mechanism through which accretion and ejection of matter take place in the protostar still remains substantially unknown from an observational point of view. Measurements of the accretion and ejection parameters are therefore of great importance in order to provide the observational constraints needed to disentangle among the proposed models.
A quantity of interest is the ratio $\dot{M}_\mathrm{loss}/\dot{M}_\mathrm{acc}$, whose expected value is in general of the order of 0.01--0.1 in disc wind models \citep[e.g.][]{ferreira97,casse00}, while a (two-sided) value of 0.3 is needed in the X-wind scenario of \citet{shu94a}. 

In our case, the measured loss rates of order $10^{-8}-10^{-7}$ M$_{\sun}\,yr^{-1}$  are comprised in the range of values that have been observed for other Class I and Class II sources \citep[e.g.][]{podio06,gullbring98,white04}, while the observed accretion rate range of $10^{-7}-10^{-6}$ M$_{\sun}\,$yr$^{-1}$ (derived from the \ion{H}{i} line luminosity) extends above the values found for the most actively accreting Class II jet sources  \citep{mohanty05}.

From these measurements we derive a mass flux that is $\sim$0.01 of the accretion rate in HH26 IRS and in HH34 IRS, which is in good agreement with values derived in T Tauri stars and with models, while for HH46 IRS a larger value $>$0.1 is inferred (see Table \ref{param1}).

Of course, the reliability of these values depend on the precision with which $\dot{M}_\mathrm{loss}$ and $\dot{M}_\mathrm{acc}$ are determined. 
These quantities might be affected for example by systematic errors due to the methods used to derive them. 
For instance, mass flux rates measured from optical or near-IR line luminosities  may be underestimated in the presence of a dense and cold jet neutral component, that is not traced by the considered lines. Similarly, 
mass accretion rates estimates are affected by the intrinsic scatter of the Muzerolle's relation and by the stellar parameters inferred assuming the source on the birthline.  

Nevertheless, the observed spread could also be explained, at least in part, in terms of variable accretion and ejection rates of the sources.
The accretion process (and consequently the related ejection of matter) is in fact likely to undergo phases of burst and quiescence \citep[see e.g.][]{larson84}, similarly to what has been suggested for other kind of accreting astronomical objects. In this scenario Class I sources would display phases of enhanced accretion  with a corresponding increase of the total luminosity.
Evidences in this sense have been provided for few Class I sources \citep[e.g. SSV13,][]{liseau92} and the existence of accretion/ejection phases is clearly observed in the jets themselves, where various distinct knots of emission are visible, often at the same distance from the source, in the red and blue lobes.

We remark that variability of the accretion can produce systematic biases when comparing the rates. This is due to the fact that the estimates of the mass flow can be done only on the jet knot closer to the protostar and therefore are derived observing matter that was ejected before the accretion process ongoing at the moment of observations, whose signatures we detect on the spectrum. 
In this case, a better estimate would take into account, rather than the \textit{instantaneous} accretion rate measured from the spectra, an accretion rate averaged over the time during which the matter responsible for the jet emission has been ejected.

Unfortunately, the values of the $\dot{M}_\mathrm{loss}/\dot{M}_\mathrm{acc}$ ratio are probably affected by this kind of problems in a significant number of cases.

\section{Conclusions}

We have investigated the accretion and ejection properties of three embedded sources (HH26 IRS, HH34 IRS, HH46 IRS) showing prominent jet-like 
structures. To this aim we have analysed their medium resolution (R $\sim$ 9000) near IR spectra acquired with VLT-ISAAC.
The main results we obtained can be summarised as follows:
\begin{itemize}

\item The bolometric luminosity and SEDs of the three sources have been revised on the basis of Spitzer and recent sub-mm observations, and theoretical radiative transfer models available in the literature. It turns out that the sources are probably very young, in a transition phase between Class 0 and I.

\item The spectra of the three sources show important differences in their characteristics: in fact, the number and the
absolute and relative intensity of the observed emission features (associated both with the accretion region and the jet environment) vary among the sources. In particular, we point out that there is no clear relationship between the presence of the jet and the spectral accretion signatures detected. Moreover, the spectra show no sign of absorption features, indicating large amount of veiling which in turn suggests the presence of warm dusty envelopes around the sources, heated by the active ongoing accretion.
\item In two of our sources (HH34 IRS and HH26 IRS) we find a  Br$\gamma$/\ion{Na}{i} ratio much larger (by a factor of three or more) than that observed in other Class Is or T Tauri stars. Conversely, the ratio between the CO(2-0) 2.3$\mu$m overtone emission and the \ion{Na}{i} 2.20$\mu$m doublet, shows the opposite trend, i.e. is smaller in the objects with jets. This may indicate the presence of massive discs around the jet sources, characterised by large amounts of warm gas in neutral state.

\item We consistently derive $A_K$, $L_*$ and $L_{\mathrm{acc}}$ assuming the three sources on the birthline and adopting the relationship between Br$\gamma$ luminosity and accretion luminosity derived by \citet{muzerolle98}. 
In the case of HH34 IRS and HH26 IRS,  accretion largely contributes to the total energy, as expected for young sources in the main accretion phase. In HH46 IRS, we estimate an $L_{\mathrm{acc}}/L_{\mathrm{bol}}$ ratio of the order of 0.2, but the real value largely depends on how the bolometric luminosity is shared among the two binary components. 
Taking into account $K$-band veiling values $r_K$ greater than the lower limits inferred from the spectra would lead in general to higher $A_K$, $L_{\mathrm{acc}}$ and $\dot{M}_{acc}$.

\item Mass accretion and loss rates span (including errors) in the range $10^{-8}$--$10^{-6}$ M$_{\sun}$\,yr$^{-1}$. The derived $\dot{M}_{loss}/\dot{M}_{acc}$ ratio is $\sim$0.01-0.03 for HH26 IRS and HH34 IRS, and $>$0.1 for HH46 IRS.  These numbers are in the range of values usually predicted by models and found in the more active classical T Tauri stars.

\item Comparing the results found in this work with other spectroscopic studies recently performed on Class I sources, we conclude that the number of Class Is actually having accretion-dominated luminosities (Accretion-Dominated Young Objects, ADYOs) could be limited.
From the properties inferred in a small sample of objects we can tentatively define some criteria to characterise such sources: ADYOs have all $K$-band veiling larger than 2 and in the majority of the cases present (in addition to \ion{H}{i}) IR features of CO and \ion{Na}{i} in emission, although these latter do not directly correlate with the accretion luminosity. 
Class Is with massive jets have high $L_{\mathrm{acc}}/L_{\mathrm{bol}}$ ratios but not all the identified ADYOs present a jet. The SEDs of our small sample of three objects, suggest that accretion-dominated sources could be in an evolutionary phase in transition between Class 0 and I.

Of course, studies of the kind presented here but carried out on larger samples of possible candidates should be performed in order to test and refine the tentative criteria that we have just mentioned.

\end{itemize}

\begin{acknowledgements}
We are grateful to the referee Sylvie Cabrit, whose insightful comments and suggestions helped us to greatly improve the paper. 
This work was partially supported by the European Community's Marie Curie Research and Training Network JETSET (Jet Simulations, Experiment and Thory) under contract MRTN-CT-2004-005592.
This publication makes use of data products from the Two Micron All Sky Survey, which is a joint project of the University of Massachusetts and the Infrared Processing and Analysis Center/California Institute of Technology, funded by the National Aeronautics and Space Administration and the National Science Foundation.
This work is based in part on observations made with the Spitzer Space Telescope, which is operated by the Jet Propulsion Laboratory, California Institute of Technology under a contract with NASA.
\end{acknowledgements} 

\bibliographystyle{aa} 
\bibliography{simone_references} 

\end{document}